\documentclass[preprint,12pt]{elsarticle}
\usepackage{multirow}
\usepackage{lineno}

 \pdfoutput=1
\newcommand{\ppbar}{\ensuremath{p\bar{p}}}

\journal{Nuclear Instruments and Methods A}

\begin{document}

\begin{frontmatter}

\title{Introduction to HOBIT, a $b$-Jet Identification Tagger at the CDF Experiment Optimized for Light Higgs Boson Searches}
\tnotetext[t1]{FERMILAB-PUB-12-118-PPD}

\author[FNAL]{J.~Freeman}
\author[FNAL]{T.~Junk}
\author[FNAL,cor1]{M.~Kirby}\ead{kirby@fnal.gov}
\author[UVA]{Y.~Oksuzian}
\author[Duke]{T.J.~Phillips}
\author[FNAL]{F.D.~Snider}
\author[Pisa]{M.~Trovato}
\author[Louvain]{J.~Vizan}
\author[LBNL]{W.M.~Yao}

\cortext[cor1]{Corresponding author}

\address[FNAL]{Fermi National Accelerator Laboratory, Batavia, IL, 60510, USA}
\address[UVA]{Univeristy of Virginia, Charlottesville, Virginia, 22906, USA}
\address[Duke]{Duke University, Durham, North Caronlina, 27708, USA}
\address[LBNL]{Ernest Orlando Lawrence Berkeley National Laboratory,\\Berkeley, California, 94720, USA}
\address[Pisa]{Isituto Nazionale di Fisica Nucleare Pisa, Scuola Normale Superiore, I-56127 Pisa, Italy}
\address[Louvain]{Universit\'e catholique de Louvain, Louvain la Neuve, B-1348, Belgium}

\begin{abstract} 
We present the development and validation of the Higgs Optimized $b$ Identification Tagger (HOBIT), a multivariate $b$-jet identification algorithm optimized for Higgs boson searches at the CDF experiment at the Fermilab Tevatron. At collider experiments, $b$ taggers allow one to distinguish particle jets containing $B$ hadrons from other jets; these algorithms have been used for many years with great success at CDF. HOBIT has been designed specifically for use in searches for light Higgs bosons decaying via $H \rightarrow b\bar{b}$. This fact combined with the extent to which HOBIT synthesizes and extends the best ideas of previous taggers makes HOBIT unique among CDF $b$-tagging algorithms. Employing feed-forward neural network architectures, HOBIT provides an output value ranging from approximately -1 (``light-jet like'') to 1 (``$b$-jet like''); this continuous output value has been tuned to provide maximum sensitivity in light Higgs boson search analyses. When tuned to the equivalent light jet rejection rate, HOBIT tags 54\% of $b$ jets in simulated 120 GeV/$c^2$ Higgs boson events compared to 39\% for SecVtx,  the most commonly used $b$ tagger at CDF. We present features of the tagger as well as its characterization in the form of $b$-jet finding efficiencies and false (light-jet) tag rates. 

\end{abstract}


\end{frontmatter}

%\tableofcontents
\linenumbers

%%%%%%%%%%%%%%%%%%%%%%%%%%%%%%%%%%%%%%%%%%%%%%%%%%%%%%%%%%%%%%%%%%
\section{Introduction}
\label{sec:intro}

At CDF, the search for a light Higgs boson has been a subject of increasing interest and focus in recent years. While there have been numerous successful $b$-jet identification algorithms (commonly referred to as ``$b$ taggers'') over the years, most have been intended for use in analyses other than searches for $H \rightarrow b\bar{b}$. Aspects of a given analysis, however, such as the optimal signal-to-background ratio, or the relative rate of non-$b$ jets originating from gluons in the data sample before tagging, can influence whether a tagger is optimal for the analysis in question. Traditional taggers have tended toward a higher purity and lower efficiency than would be ideal for Higgs boson searches given the relatively low cross section of Higgs boson production at Tevatron energies. While this problem has been circumvented somewhat by taking the logical OR of several taggers, a more elegant and flexible solution can be found in the continuous output of a neural network, tunable for each analysis application. 

In this paper, we describe the Higgs Optimized $b$ Identification
     Tagger (HOBIT). The strategy used in developing HOBIT is to build upon the strengths of previous CDF $b$ taggers, address their weaknesses,
     and construct a new tagger that is highly optimized specifically for
     finding light Higgs boson decays. HOBIT produces a continuous output
     variable, allowing efficiency and background rejection to be tuned to
     meet the requirements of a given search. In the next section, we
     review some of the general features of $b$ quark decays used by HOBIT to
     distinguish jets containing $B$ hadrons from jets produced by gluons or light quarks (up, down, or strange). Section \ref{sec:algorithms} then
     describes some of the previous $b$-tagging algorithms used by CDF upon
     which HOBIT is built. We then discuss some features of the CDF
     detector in Sec.~\ref{sec:detector}, followed by a detailed description of the
     HOBIT algorithm and training regimen. The performance of HOBIT as
     characterized by the $b$-jet tagging efficiency and background rejection
     rates in data and Monte Carlo (MC) is presented in Sec.~\ref{sec:scale_factors}. We
     conclude in Sec.~\ref{sec:conclusion}.

\section{Physics of $b$'s from Higgs Boson Decay}
\label{sec:physics}

Jets containing high-$E_T$ $B$ hadrons such as are created in a light Higgs boson decay possess several features that distinguish them from jets produced by light quarks or gluons.  The most important of these is the relatively long lifetime of a $B$ hadron, augmented in the lab frame by its relativistic boost, which allows it to travel a distance on the order of a millimeter\footnote{%
This distance is achieved due to the fact that $c$ times the rest frame lifetime of a $B^{0}$ ($B^{\pm}$, $B_s$, $\Lambda_{b}$) hadron is
460 $\mu$m (501 $\mu$m, 441 $\mu$m, 367 $\mu$m).
}. The $B$ hadron's travel across these macroscopic distances results in a displacement between the location of the $p\bar{p}$ collision (the ``primary'' vertex)
and the $B$ hadron decay (the ``secondary'', or ``displaced'' vertex). These displacements are resolvable by the CDF
tracking system, and in particular by its silicon detector. Almost all
information as to whether or not a given jet originates from $b$-quark production is carried in the tracks reconstructed from detector signals left by the jet's charged particles. Specifically, it is possible to identify the decay of a $B$ hadron through the displacement from the primary vertex of the individual tracks it leaves in the detector, and also through the displacement of a $B$-hadron decay vertex formed by combining multiple displaced tracks in a fit. 

Other features also distinguish the $b$ jet from other jets. Due to the large mass of the $b$ quark, the collective invariant mass of the decay products of $B$ hadrons will be larger than those from the decay products of hadrons not containing $b$ quarks. Furthermore, the large relativistic
boost typical of a $B$ hadron will result in decay products which tend to be more
energetic and collimated within a jet cone than other particles. Finally, particle
multiplicities tend to be different for jets containing $B$ hadron
decays compared to other jets; in particular, muons or electrons
appear in approximately 20\% of jets containing a $B$ hadron, either directly via semileptonic decay of the $B$ or
indirectly through the semileptonic decay of charm hadrons resulting from a $B$ decay.

\section{$b$-Tagging Algorithms}
\label{sec:algorithms}

% J. Freeman, 1/18/12 -- not sure whether or not I want to use some parts of the old intro to the Bness NIM's description of earlier tagging algorithms, so I leave it included in the file but commented out. 

%% Many algorithms used at CDF were instrumental in the 1995 discovery of
%% the top quark~\cite{cdftop_evidence}. Here we review the standard
%% $b$-tagging algorithms used at CDF.  Similar techniques as those
%% described in this paper have been developed at the D0
%% experiment~\cite{d0tagging} and at the CMS and ATLAS experiments at
%% the LHC~\cite{CMS-PAS-BTV-11-001,ATLAS-CONF-2011-102}.

As a tremendous amount of effort has gone into the construction of $b$ taggers at CDF and other experiments~\cite{D0tag,CMStag,ATLAStag}, we build upon previous experience when constructing HOBIT. In particular, HOBIT explicitly uses as inputs the output of the SecVtx algorithm set to its ``loose'' operating point~\cite{SecVtx}, the output of CDF's soft muon tagger~\cite{SLT}, and inputs to the earlier RomaNN~\cite{roma,roma2} and Bness~\cite{bness} multivariate taggers. Consequently, it is useful to describe these taggers.

\subsection{SecVtx}

SecVtx is a displaced vertex tagger and the most
commonly used $b$ tagger at CDF. SecVtx only uses tracks which are significantly displaced from the primary vertex, accepted by quality requirements, and within a distance of $\Delta R < 0.4$ of the jet axis. Here, $\Delta R = \sqrt{ \Delta \phi^2 + \Delta \eta^2 }$, where $\phi$ is the azimuthal angle of the 
   track around the beam axis, 
   and $\eta$ is its pseudorapidity defined as $\eta = -\log (\tan(
   \frac{\theta}{2}))$, with $\theta$ the polar angle of the track with
   respect to the beam axis. With these tracks, SecVtx uses an iterative method to fit a displaced vertex within the jet, where the $\chi^2$ of the vertex fit is employed to guide the process. 
Assuming that this displacement is due to the long lifetime of the $B$ hadron, the significance
of the two-dimensional decay length $L_{xy}$ in the plane perpendicular to the beampipe axis is used to select $b$-jet candidates. The algorithm is utilized with different track requirements and threshold
values in order to achieve different efficiencies and purity rates. In practice, three operating points are used, referred to as
``loose'', ``tight'', and ``ultra-tight''. The loose SecVtx operating point decision is used as an input to both the RomaNN and HOBIT tagger. One drawback of the SecVtx tagger is that it is unable to fit a vertex in every $b$ jet. In the Pythia~\cite{pythia} 120 GeV/$c^2$ Higgs boson Monte Carlo (MC) whose $b$ jets are used to train HOBIT, SecVtx operating at its ``loose'' setting fails to find a vertex in 44.3\% of these jets. 

\subsection{Soft Lepton Taggers}

Soft lepton taggers~\cite{SLT} (SLT) take a different approach to $b$
tagging. Rather than focusing on tracks within a jet, they select $B$ hadron decays by identifying charged leptons inside a cone around the jet axis. Since the $b$ semileptonic branching ratio is approximately 10\% per lepton flavor, this class of tagger is not competitive with SecVtx or the other taggers described below if used alone. However, because a soft lepton tagger does not rely on the presence of displaced tracks or vertices, it has a chance to identify $b$ jets that the other methods cannot. In practice, CDF uses only a soft muon tagger since high-purity electron or $\tau$ identification within jets is difficult. HOBIT uses as inputs the number of soft muon tags within a jet as well as the momentum transverse to the jet axis of the muon with the highest-likelihood tag.

%The soft muon tagger is used to identify muons within a jet, whose multiplicities and momenta transverse to the jet axis were used as inputs to RomaNN, and are used now as inputs to HOBIT.

\subsection{The RomaNN Tagger}

The ``RomaNN tagger" has been used at CDF in light Higgs boson searches~\cite{roma,roma2} and employs neural network architectures. Neural networks (NNs) can use as many flavor-discriminating
observables as is computationally feasible; hence the efficiency of NN
taggers is equal to or greater than that of conventional taggers for a
given purity. While the SecVtx tagger
attempts to find exactly one displaced vertex in a jet, the RomaNN
tagger uses a vertexing algorithm that can find multiple vertices, as
may be the case when multiple hadrons decay within the same jet cone
(for example, in a $B \rightarrow D$ decay). The RomaNN tagger uses several types
of NNs: one to distinguish vertices which come from a heavy flavor ($B$ or charm) hadron from false vertices or vertices coming from other hadrons; another to identify unvertexed tracks
which come from a heavy flavor hadron; and then another NN which takes as inputs the output of the first NNs along with other inputs, including the loose SecVtx tag status, the number of
SLT-identified muons, and the vertex displacement and mass
information. Distinct versions of this third NN are trained to separate $b$ jets from light jets, charm jets from light jets, and $b$ jets from charm jets; the outputs of these three flavor-separating NNs are then used to train a final NN whose output is the RomaNN discrimination variable. The RomaNN tagger not only has superior performance to
that of SecVtx at equivalent purities (see Fig.~\ref{roc_comp} ), but
also allows for an ``ultra-loose'' operating point yielding greater
efficiency, particularly useful in light Higgs boson searches. 

However, the RomaNN tagger is not guaranteed to fit a vertex or to have sufficient
input information to reliably tag a jet. In the event that the RomaNN tagger fails to receive sufficient information from its inputs, it is unable to assign an output value to that jet. This is the case with 20.6\% of the $b$ jets in the aforementioned light Higgs boson MC sample. Regardless, due to the usefulness of the RomaNN inputs, a majority of them are employed as inputs into the HOBIT tagger, which allows HOBIT to take advantage of the same extensive vertex information that the RomaNN tagger uses.

%. This allows HOBIT to take advantage of the same extensive vertex information of which RomaNN takes advantage.

\subsection{The Bness Tagger}

While the RomaNN tagger focuses on the vertices it finds within a jet, in the event that it is unable to fit any
vertices, it is unable to distinguish $b$ jets from light
jets. However, a significant proportion of $b$ jets (approximately 20\% in Higgs boson candidate events) do not contain a sufficient number of well-reconstructed tracks to allow for
a vertex fit in the RomaNN tagger. The Bness tagger~\cite{bness} uses not only vertex information within a jet, but also the properties of individual tracks to determine whether a jet is $b$-like. (The RomaNN tagger only examines individual tracks based on their proximity to a displaced vertex). To evaluate the information from individual tracks, the Bness tagger utilizes an NN which is applied to all tracks passing loose requirements, and which takes positional
(e.g., impact parameter) and kinematic (e.g., $p_T$) information on a
track to determine whether it appears to have come from the decay of a
$B$ hadron. The Bness tagger is therefore able to extract information from all but a few percent of $B$ jets,
and can achieve a very high efficiency for a reasonable level of
purity. This robust property of the tagger makes it useful for
analyses where efficiency is critical,
as is the case with light Higgs boson analyses or even searches for hadronic
decays of heavy gauge bosons (see Ref.~\cite{WZ} for more details). A track-by-track NN very similar to that employed by the Bness tagger is
used to evaluate tracks in HOBIT; this will be described in Section~\ref{sec:hobittagger}. One drawback of the Bness tagger is that, like SecVtx and unlike RomaNN, it is only able to fit one vertex per jet. Additionally, it uses fewer vertex-based inputs than the RomaNN tagger, and therefore only its track-by-track algorithm is used in HOBIT.

%%%%%%%%%%%%%%%%%%%%%%%%%%%%%%%%%%%%%%%%%%%%%%%%%%%%%%%%%%%%%%%%%%%%%%%%%%%%%%%%

%CONSIDER MOVING SECTION ON CDF DETECTOR TO BEFORE THE DESCRIPTION OF
%THE B TAGGING ALGORITHMS

\section{The CDF Detector}
\label{sec:detector}

The CDF~II detector is described in detail
elsewhere~\cite{CDF_detect_A}. The detector is cylindrically
symmetric around the proton beam line\footnote{%
  The proton beam direction is defined as the positive $z$
  direction. The rectangular coordinates $x$
  and $y$ point radially outward and vertically upward from the Tevatron
  ring, respectively. Transverse energy, and transverse momentum are defined as $E_{T}$=$E\sin\theta$, and
  $p_{T}$=$p\sin\theta$, respectively, $\theta$ having been defined in Sec.~\ref{sec:algorithms}}
%% Pseudorapidity,
%%   transverse energy, and transverse momentum are defined as
%%   $\eta$=$-\ln\tan(\theta/2)$, $E_{T}$=$E\sin\theta$, and
%%   $p_{T}$=$p\sin\theta$, respectively.  The rectangular coordinates $x$
%%   and $y$ point radially outward and vertically upward from the Tevatron
%%   ring, respectively.}
%
with tracking systems that sit within a superconducting solenoid which
produces a $1.4$~T magnetic field aligned coaxially with the $\ppbar$
beams. A set of calorimeters and muon detectors, to be described later, surround the tracking systems and solenoid. 

The outermost tracking system, the Central Outer Tracker (COT), is a $3.1$ m long open cell
drift chamber which performs up to 96 track position measurements in the region
between $0.40$ and $1.37$ m from the beam axis, providing coverage in
the pseudorapidity region $|\eta| \le 1.0$~\cite{cot_nim} . Sense wires
are arranged in eight alternating axial and $\pm2^{\circ}$ stereo
``superlayers" with 12 wires each. The position resolution of a single
drift time measurement is about $140~\mu$m.

Charged-particle trajectories are found first as a series of
approximate line segments in the individual axial superlayers. Two
complementary algorithms associate segments lying on a common circle,
and the results are merged to form a final set of axial tracks. Track
segments in stereo superlayers are associated with the axial track
segments to reconstruct tracks in three dimensions. 

%% probably not needed for this NIM
% A road-based hardware pattern recognition algorithm runs online in the eXtremely Fast Tracker (XFT) to provide track information for triggering. Drift times partitioned into two time bins are used to find axial segments which are matched in their positions and slopes. An ``XFT" track is one which has four matching axial segments along a trajectory. The XFT efficiency is measured in a set of well-measured COT tracks passing all four axial superlayers, and is found to be $96.7 \pm 0.1 \%$ for charged particles with $p > 25$ GeV/$c$.

A five layer double-sided silicon microstrip detector (SVX) covers
the region between $2.5$ to $11$ cm from the beam axis. Three separate
SVX barrel modules along the beam line together cover a
length of 96 cm, approximately 90\% of the luminous beam interaction
region. Three of the five layers combine an $r$-$\phi$ measurement on
one side and a $90^{\circ}$ stereo measurement on the other, and the
remaining two layers combine an $r$-$\phi$ measurement with small angle stereo at
$\pm1.2^{\circ}$. The typical silicon hit resolution is 11~$\mu$m. Additional Intermediate Silicon Layers (ISL) at radii between 19
and 30 cm from the beam line in the central region link tracks in the
COT to hits in the SVX.

Silicon hit information is added to COT tracks using a
progressive ``outside-in" tracking algorithm in which COT tracks are
extrapolated into the silicon detector, associated silicon hits are
found, and the track is refit with the added information of the
silicon measurements. The initial track parameters provide a width for
a search road in a given layer. Then, for each candidate hit in that
layer, the track is refit and used to define the search road into the
next layer. This stepwise addition of precision SVX information at
each layer progressively reduces the size of the search road, while
also accounting for the additional uncertainty due to multiple
scattering in each layer. The search uses all candidate hits
in each layer to generate a small tree of final track candidates, from
which the tracks with the best $\chi^{2}$ are selected. The efficiency
for associating at least three silicon hits with an isolated COT track
is $91 \pm 1\%$. The extrapolated impact parameter resolution for
high-momentum outside-in tracks is much smaller than for COT-only
tracks: $40~\mu$m, dominated by a $30~\mu$m uncertainty in the beam position.

Outside the tracking systems and the solenoid, segmented calorimeters
with projective geometry are used to reconstruct electromagnetic (EM)
showers and jets. The EM and hadronic calorimeters are
lead-scintillator and iron-scintillator sampling devices,
respectively. The central and plug calorimeters are segmented into
towers, each covering a small range of pseudorapidity and azimuth,
and in full cover the entire $2\pi$ in azimuth and the pseudorapidity
regions of $|\mathrm\eta|$$<$1.1 and 1.1$<$$|\mathrm\eta|$$<$3.6
respectively. The transverse energy, $E_T$, where the
polar angle is calculated using the measured $z$ position of the event
vertex, is measured in each calorimeter tower. Proportional chambers and
scintillation detectors arranged in strips measure the transverse profile of EM
showers at a depth corresponding to the shower maximum.

High-momentum jets, photons, and electrons leave isolated energy
deposits in contiguous groups of calorimeter towers which can be
summed together into an energy ``cluster''. 
%For the purpose of
                                %triggering, online processors
                                %organize the calorimeter tower
                                %information into separate lists of
                                %clusters for the electromagnetic
                                %compartments alone, and for the
                                %electromagnetic and hadronic
                                %compartments combines. 
Electrons are identified in the central EM calorimeter as isolated,
mostly electromagnetic clusters that also match with a track in the
pseudorapidity range $|\eta| < 1.1$. The electron transverse energy is
reconstructed from the measured energy in the electromagnetic cluster with precision
$\sigma(E_{T})/E_{T} = 13.5\% / \sqrt{E_{T} ({\rm GeV})} \oplus 2\%$,
where the $\oplus$ symbol denotes addition in quadrature. Jets are
identified as a group of electromagnetic and hadronic calorimeter
clusters using the \textsc{jetclu} algorithm~\cite{jetclu} with a cone
size of $\Delta R = 0.4$. Jet energies are corrected for calorimeter non-linearity, losses in the gaps betwen towers, multiple primary
interactions, the underlying event, and out-of-cone
losses~\cite{jesnim} . The jet energy resolution is approximately
$\sigma_{E_{T}} = 1.0~{\rm GeV} + 0.1\times E_T$.

Directly outside of the calorimeter, four-layer stacks of planar drift
chambers detect muons with $p_{T} > 1.4~$GeV$/c$ that traverse 
the five absorption lengths of the calorimeter. Farther out, behind an
additional 60 cm of steel, four layers of drift chambers detect muons
with $p_{T} > 2.0~$GeV$/c$. The two systems both cover the region $|\eta| \le 0.6$, though they have different structures, and therefore places where the geometrical coverage does not overlap. Muons in the region $0.6 \le |\eta| \le 1.0$ pass through at least four drift
layers arranged in a conic section outside of the central
calorimeter. Muons are identified as isolated tracks in the COT that
extrapolate to track segments in one of the four-layer stacks.

\section{ The HOBIT Tagger}
\label{sec:hobittagger}

%%%%%%%%%%%%%%%%%%%%%%%%%%%%%%%%%%%%%%%
%BEGIN HOBIT SECTION
%%%%%%%%%%%%%%%%%%%%%%%%%%%%%%%%%%%%%%%

%% The HOBIT tagger possesses all the advantages of other multivariate
%% taggers such as RomaNN~\cite{roma,roma2} and Bness~\cite{bness} , most noteably a near-maximal
%% use of the information available in a $b$ jet and a tunable purity-efficiency

The HOBIT tagger is similar to other multivariate $b$-tagging algorithms previously used at CDF, such as the RomaNN and Bness
	   taggers. All of these taggers attempt to make maximal use of the
	   available information in $b$ jets, and construct a continuous
	   discriminating variable. HOBIT improves upon these earlier
	   taggers, however, by addressing specific weaknesses of each and
	   optimizing for light Higgs boson searches.

\subsection{The architecture}

HOBIT is constructed as a feed-forward multilayer perceptron neural
network implemented using the TMVA package for Root~\cite{TMVA}. It
consists of two hidden layers of 25 and 26 nodes, there being 25 inputs to the tagger, and a hyperbolic tangent activation function. Five hundred cycles were used in the training. The training regimen used $b$ jets in Pythia~\cite{pythia} 120 GeV/$c^2$ Higgs boson Monte Carlo (MC) and
light jets from Alpgen-generated Pythia $W+$jets MC. Charm jets were not considered during training due to preliminary studies which indicated a relative
insensitivity of light Higgs boson searches to charm jet
contamination. Here, ``$b$ jet'' denotes a jet with a $B$ hadron
within a cone of $\Delta R < 0.4$ of the jet axis, while a ``charm jet''
contains a charm hadron but no $B$ hadrons within this cone and a
``light jet'' contains neither $B$ hadrons nor charm hadrons within this cone. Jets were required to have an $E_T > 15$ GeV, $|\eta| < 2$, and
at least one track for use in the track-by-track NN described in Sec.~\ref{sec:hobittagger:track-by-track}.

The 25 inputs to the tagger are a combination of RomaNN and Bness inputs, albeit with some exceptions, additions and modifications. Fourteen of these inputs are also inputs to the RomaNN tagger. A further ten inputs to HOBIT are the ten highest track-by-track NN discriminant output values of tracks in the jet cone. In the event that there are fewer than ten tracks in a jet, the value of the remaining track-by-track NN inputs are set to -1 as this is the light-jet-like value of the NN output. The number of tracks which pass the track-by-track NN selection criteria is found to have additional discriminating power and is also used as an input to HOBIT. Track selections differ between tracks used for RomaNN inputs and tracks evaluated with the track-by-track NN. Tracks used for RomaNN inputs must have $p_T > 1$ GeV/c and be within $\Delta R < 0.4$ of the jet axis (the same selection used in the published RomaNN tagger), while tracks used by the track-by-track NN inputs had a looser requirement of $p_T > 0.5$ GeV/c and a distance of $\Delta R < 0.7$ from the jet axis (the original requirement was $\Delta R < 0.4$). Other selection cuts were considered, but none resulted in an improvement in the performance of HOBIT. Note that one of the RomaNN inputs used (also used in the Bness tagger) is the $E_T$ of the jet itself. The various HOBIT inputs are correlated with $E_T$, so the $E_T$ provides additional useful information to HOBIT. We prevent kinematic biasing of HOBIT by weighting the light jet training sample to have the same $E_T$ distribution as the $b$-jet training sample.

As previously mentioned, one potential weakness of the RomaNN tagger is its inability to produce a useable output when there is insufficient input information. This requirement of ``RomaNN taggability''
can be a liability when very high $b$-jet tagging efficiency is sought. In the MC sample used to train the HOBIT tagger, 21\% of $b$ jets fail to be RomaNN taggable, versus 30\% of light
jets. The track-by-track NN in HOBIT compensates for this shortfall of RomaNN. While jets in HOBIT are required to have at least one track with an evaluated track-by-track NN output, only 3.0\% of $b$ jets and 2.1\% of light jets in the MC fail this requirement, indicating a very efficient taggability requirement.  

The full list of inputs to HOBIT ranked by importance after TMVA's training is provided in Table~\ref{tab:hobit_inputs}. Here, ``importance'' refers to the sum of the squares of the
weights connecting a given input to the nodes of the first hidden layer of HOBIT. Distributions of the inputs to HOBIT are shown in Fig.~\ref{hobit_inputs}. A description of these inputs is given below. 

\begin{table}[hbt]
\begin{center}
%\begin{tabular}{|cc|cc|} \hline
\begin{tabular}{|c|c|c|} \hline
Jet (HOBIT) Input & Importance \\ \hline
RomaVtx pseudo-c$\tau$ & 435 \\
RomaVtx 3-d displacement significance & 382 \\
Bness 0 & 77.5 \\
Bness 1 & 21.5 \\
SecVtx Loose & 16.9 \\
Bness 3 & 9.90 \\
Number of muons & 7.80 \\
ptFrac & 7.05 \\
Bness 2 & 6.22 \\
Bness 4 & 5.46 \\
muon $p_T$ to jet axis & 5.32 \\
Bness 5 & 4.54 \\
Bness 9 & 4.46 \\
M$_{inv}$ of HF-like tracks & 4.17 \\
Bness 6 & 3.44 \\
Bness 8 & 2.70 \\
RomaVtx 3-d displacement & 2.24 \\
SecVtx Mass & 1.68 \\
Bness 7 & 1.51 \\
RomaVtx Mass & 0.752 \\
Number of track-by-track NN tracks & 0.380 \\
Number of HF-like tracks & 0.287 \\
Jet $E_T$ & 0.161 \\
Number of Roma-selected tracks & 0.125 \\
Total $p_T$ of tracks & 0.00250 \\
\hline
\end{tabular}
\caption{Inputs to the HOBIT tagger and their importances;
        ranking is done by importance (see text for definition of this term). ``RomaVtx'' denotes the most HF-like vertex as found by the RomaNN tagger. }
\label{tab:hobit_inputs}
\end{center}
\end{table}

\subsection{The RomaNN inputs}

RomaNN inputs used in HOBIT consist of observables built using tracks
and vertices found to be ``heavy-flavor-like'' (HF-like) according to its NNs. No modifications were made to the RomaNN inputs compared to the published tagger. These inputs include:

%including the properties of the fitted vertex found to be the most heavy-flavor-like (its displacement, invariant track mass, pseudo-c$\tau$), as well as the number of SLT-tagged muons and the jet's loose SecVtx tagger status. Note that ``heavy-flavor'' (HF) denotes not only $B$ hadrons but those which contain charm as well, which also have relatively long lifetimes. 

\begin{itemize}

\item The invariant mass, pseudo-c$\tau$, 3-d displacement and 3-d
 displacement significance of the most HF-like vertex.

%\item The 3-d displacement of the 2nd most HF-like vertex.

\item The number of tracks both in HF-like vertices and standalone HF-like
 tracks associated to a displaced vertex, as well as their combined invariant mass, and the ratio of the scalar sum of the $p_T$'s of
 these tracks to the scalar sum of the $p_T$'s of all tracks in the jet.

\item The loose SecVtx tag status, as well as the mass of the tracks used in the loose SecVtx vertex fit.

\end{itemize}

\subsection{Bness inputs: the track-by-track NN}
\label{sec:hobittagger:track-by-track}

As mentioned above, the ten highest evaluated track-by-track NN outputs for tracks in a jet serve as inputs to HOBIT. Therefore, this section concerns the track-by-track NN itself. The input variables to the track-by-track NN are the same for HOBIT as were used in the track-by-track NN of the original Bness tagger. However, the track-by-track Bness NN was retrained to create the HOBIT track-by-track NN. This was done not only because the cone requirement on the tracks was loosened but also because we wished to optimize the track-by-track NN for light Higgs boson searches. Hence, while the original Bness track-by-track NN was trained using $ZZ \rightarrow$ 4 jets MC, the HOBIT track-by-track NN was trained using the same MC as was used to train the overall HOBIT tagger. Since the track-by-track NN operates at the level of individual tracks, we impose an additional requirement on $b$-jet tracks for the purposes of training by demanding that they be within $\Delta R < 0.05$ of the actual charged particles resulting from a $B$ hadron decay in the MC. The track-by-track NN employed the same basic framework for training as that used for HOBIT itself (training cycles, inner layer structure, etc.).

\begin{figure}[htbp]
  \begin{center}
    \includegraphics[width=0.9\textwidth,clip=]{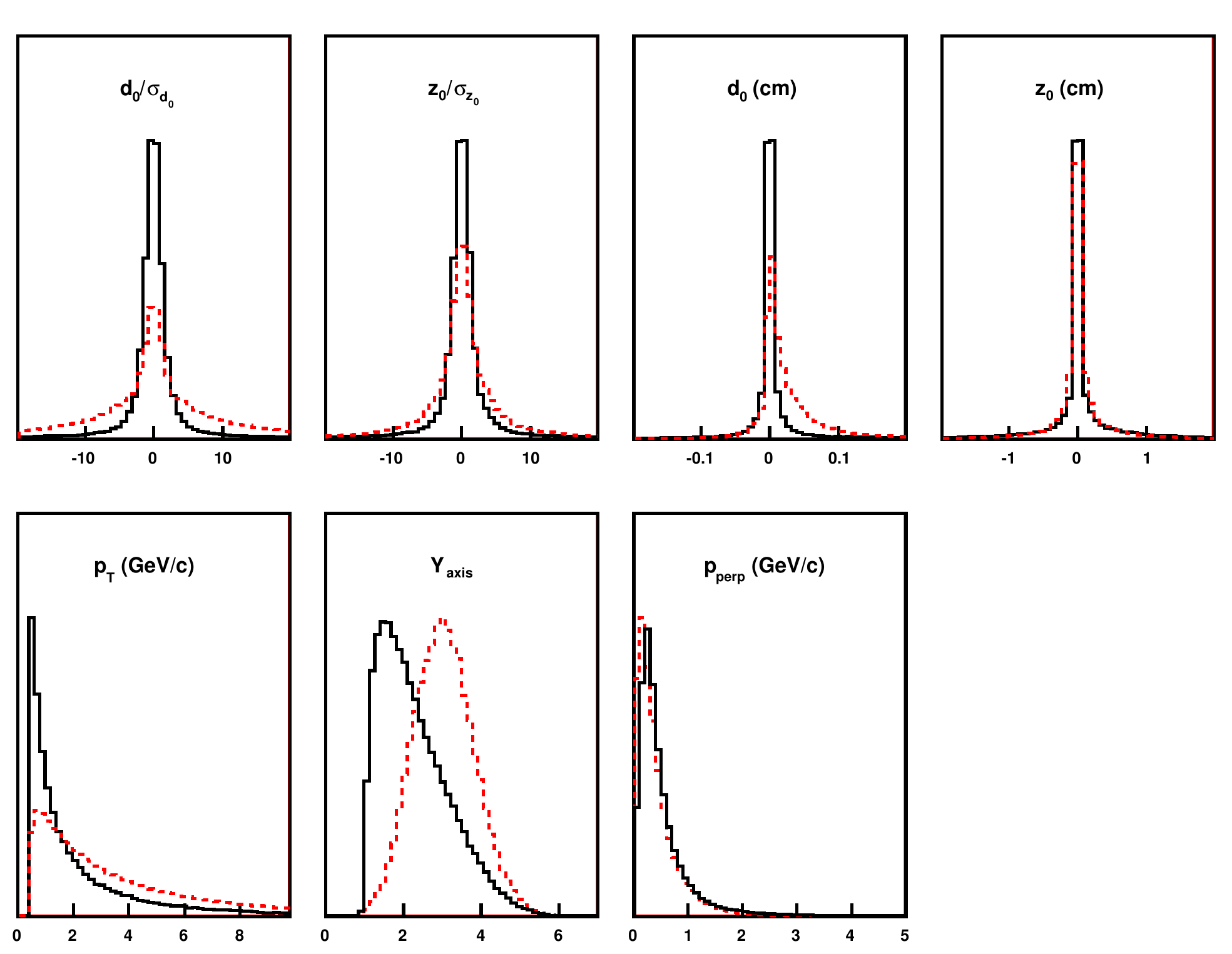}
    \caption{Inputs to track-by-track NN. The solid histogram is for tracks in light quark jets and the dashed (colored) histogram is for tracks in $b$ jets; taken from MC, the distributions are normalized to one another. Not shown is the jet $E_T$, identical between the two distributions by construction. Left-to-right, top-to-bottom: significance of the impact parameter and $\Delta z$ between the track and the primary vertex; the values of the impact parameter and $\Delta z$; the $p_T$ of the track with respect to the beam axis; and the track's rapidity and $p_T$ with respect to the jet axis. \label{track_inputs}}
  \end{center}
\end{figure}

Some of the inputs to the track-by-track NN take advantage of the fact that tracks from $B$ hadron decays are displaced from the primary vertex. These
    inputs include the impact parameter, the distance along the z-axis between
      the track and the primary vertex, and the significance of each. Kinematic inputs such as the $p_T$, rapidity, and track momentum perpendicular to the jet axis ($p_\textrm{perp}$) exploit the greater collimation of $B$ tracks due to the large boost of the hadron. Finally, the jet $E_{T}$ is an input to the track-by-track NN, because the previously mentioned inputs are correlated with jet $E_{T}$. Tracks from light jets are weighted in training such that the jets which contain them have the same $E_T$ distribution as the $b$ jets; this is done so as to avoid kinematic biasing in the track-by-track NN. Distributions of the track-by-track NN inputs are shown in Fig.~\ref{track_inputs}. Not shown are the jet $E_T$ distributions, which are identical by construction.

\begin{figure}[tbp]
  \begin{center}
    \includegraphics[width=0.9\textwidth,clip=]{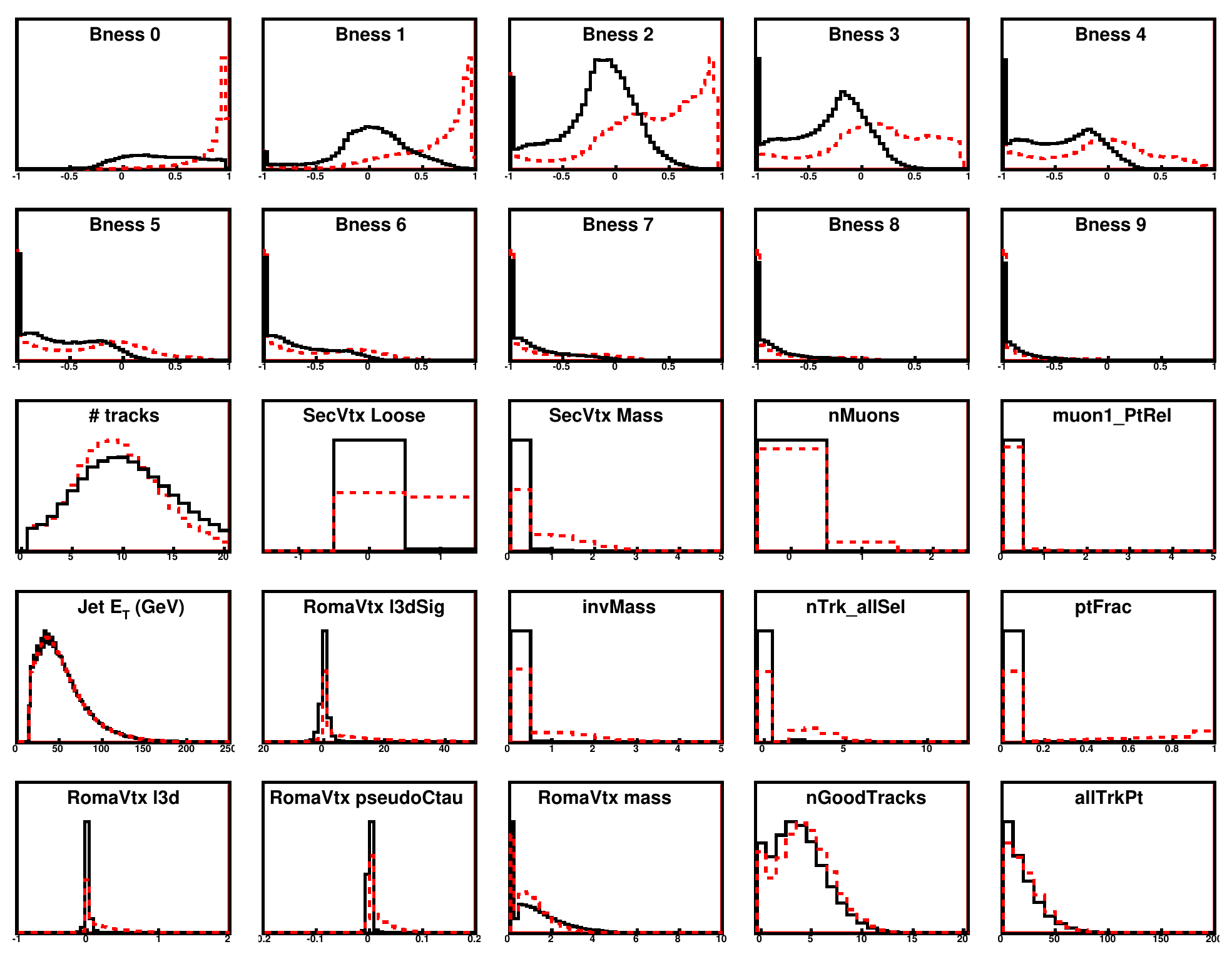}
    \caption{Inputs to HOBIT. The solid histogram is for light quark jets and the dashed (colored) histogram is for $b$ jets. Taken from MC, the distributions are normalized to one another. Left to right, top to bottom: the Bness value for the 10 highest Bness tracks; the number of Bness-selected tracks; the loose SecVtx tag status and the mass of its fitted vertex; the number of SLT-tagged muons and the momentum transverse to the jet axis of the most SLT-favored muon; jet $E_T$; the 3-d displacement significance of the most HF-like vertex in RomaNN; the invariant mass, number, and fraction of total track $p_T$ of HF-like tracks; the 3-d displacement, pseudo-c$\tau$ and invariant mass of the most HF-like vertex; the number of RomaNN-selected tracks and their total $p_T$.
             \label{hobit_inputs}}
  \end{center}
\end{figure}

\subsection{HOBIT Performance}

\begin{figure}[htbp]
  \begin{center}
    \includegraphics[width=0.9\textwidth,clip=]{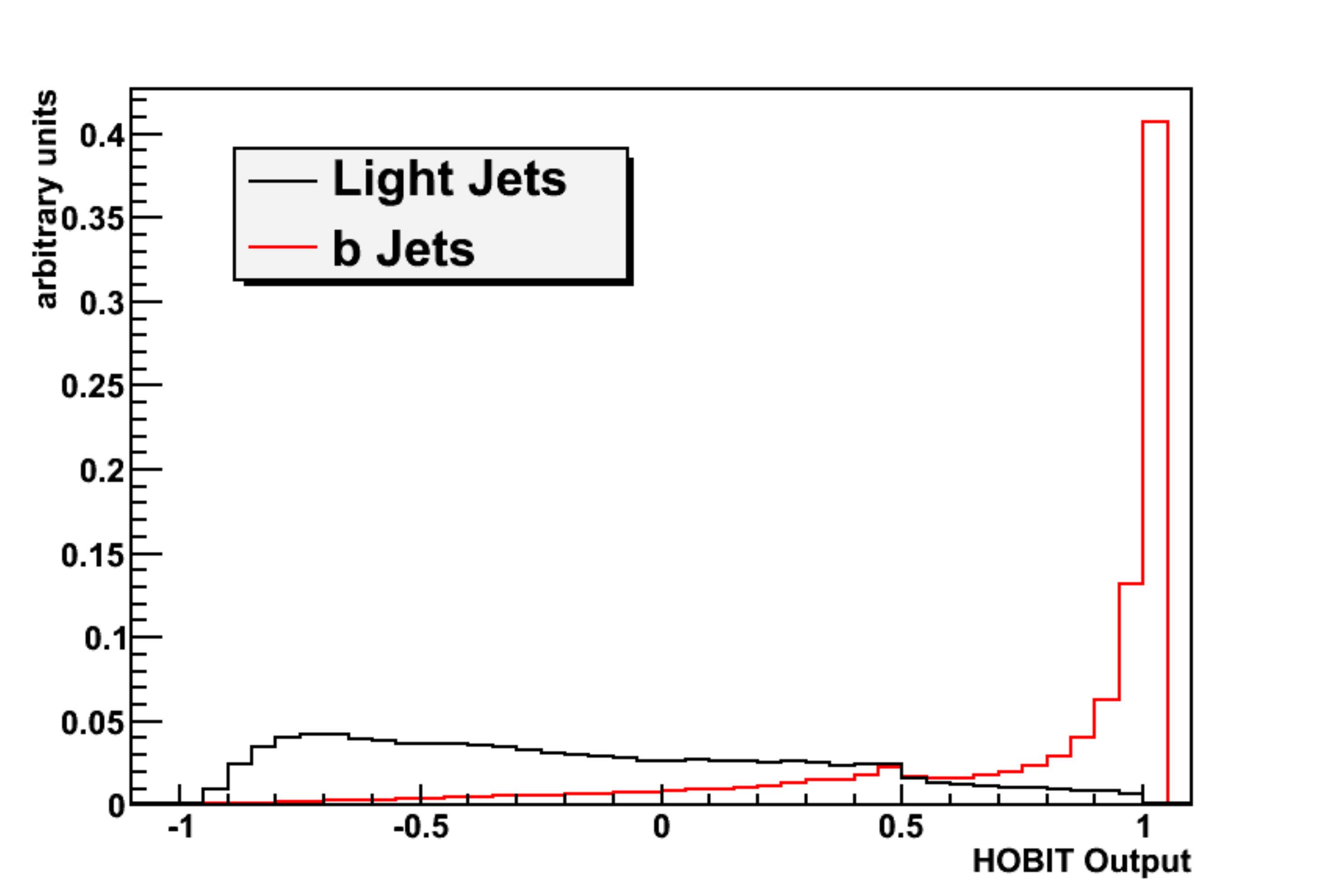}
    \caption{HOBIT outputs. The output is trained so that 1 is $b$ jet-like and -1 is targeted to be light jet-like. The black histogram is for light quark jets and the colored histogram is for $b$ jets. Taken from MC, the distributions are normalized to one another.
             \label{hobit_outputs}}
  \end{center}
\end{figure}

\begin{figure}[htbp]
  \begin{center}
    \includegraphics[width=0.9\textwidth,clip=]{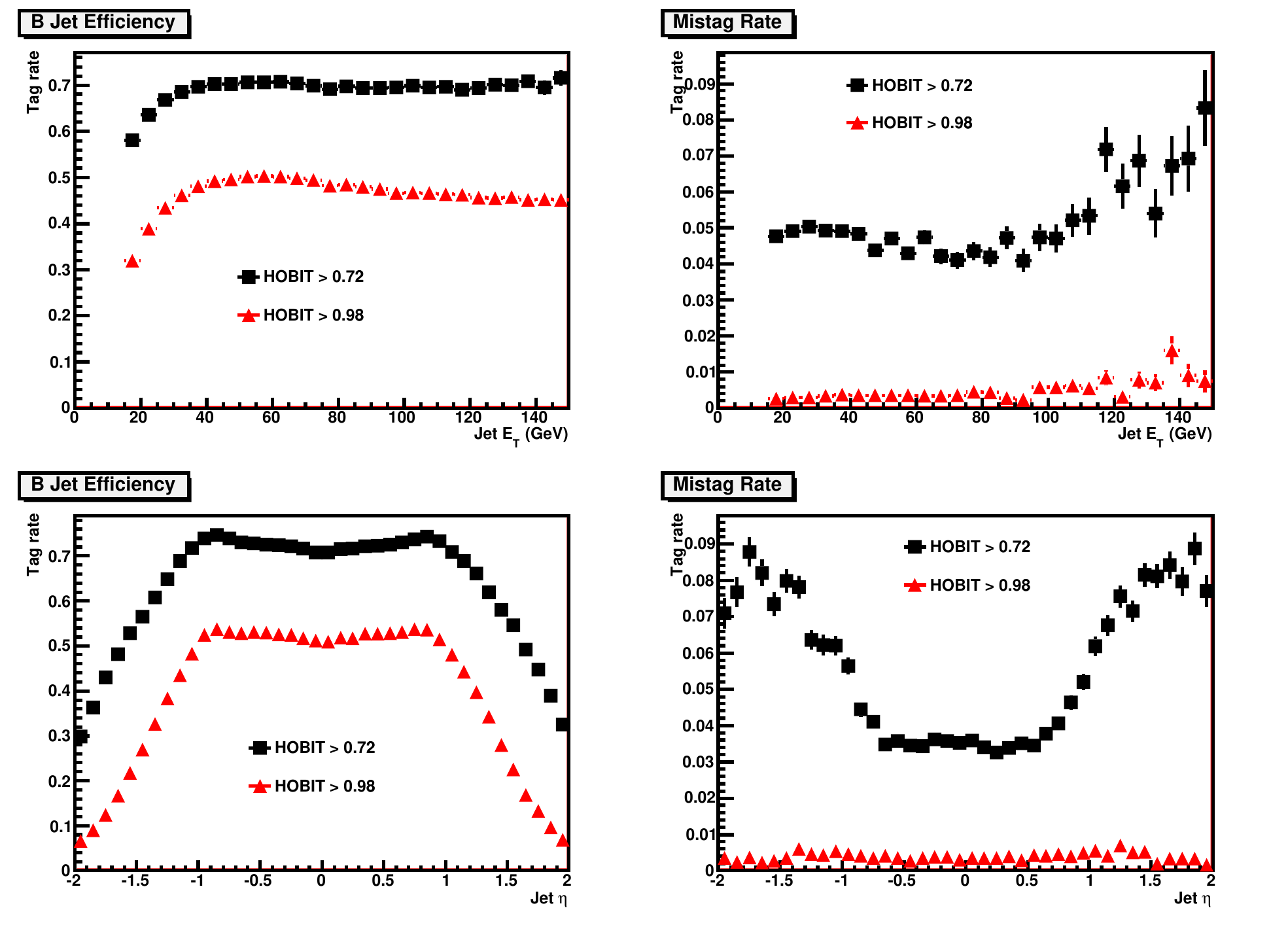}
    \caption{The $b$-jet and light-jet efficiencies in MC before SF corrections as a function of $\eta$ and $E_{T}$.  The black triangles are for the looser operating point and the colored triangles are for the tighter operating point. 
             \label{eff_vs_kinematic}}
  \end{center}
\end{figure}

 The output HOBIT distributions for $b$-jets and light-jets from an independent but identically generated MC sample as was used to train the discriminator are shown in Fig.~\ref{hobit_outputs}. In Fig.~\ref{eff_vs_kinematic}, the $b$-jet efficiencies and the light jet efficiencies (``mistag rates'') as a function of jet $E_{T}$ and $\eta$ are shown for two HOBIT operating points -- a requirement of a HOBIT output $>$ 0.72 (``loose'') and a requirement of a HOBIT output $>$ 0.98 (``tight''). At higher $\eta$, where tracking coverage is more sparse and less information is available, the $b$-tagging efficiency drops, as would be expected. Interestingly, the mistag rate increases in the case of the loose tag and drops in the case of the tight tag, demonstrating the higher impact of incorrectly identified tracks when using a loose tagging requirement. In general, the efficiency increases with increasing jet $E_{T}$ due to the greater displacement of the $B$ hadron. Similarly, the light jet efficiency increases, at least in part due to the higher rapidity and $p_T$ of tracks in high-$E_T$ jets.  

%on the position measurements of higher-$p_T$ tracks and also due to actual long lived particles in light jets (such as $K_{s}$ and $\Lambda$) having a greater displacement and thus being misidentified as $B$ hadrons.

\begin{figure}[htbp]
  \begin{center}
    \includegraphics[width=0.9\textwidth,clip=]{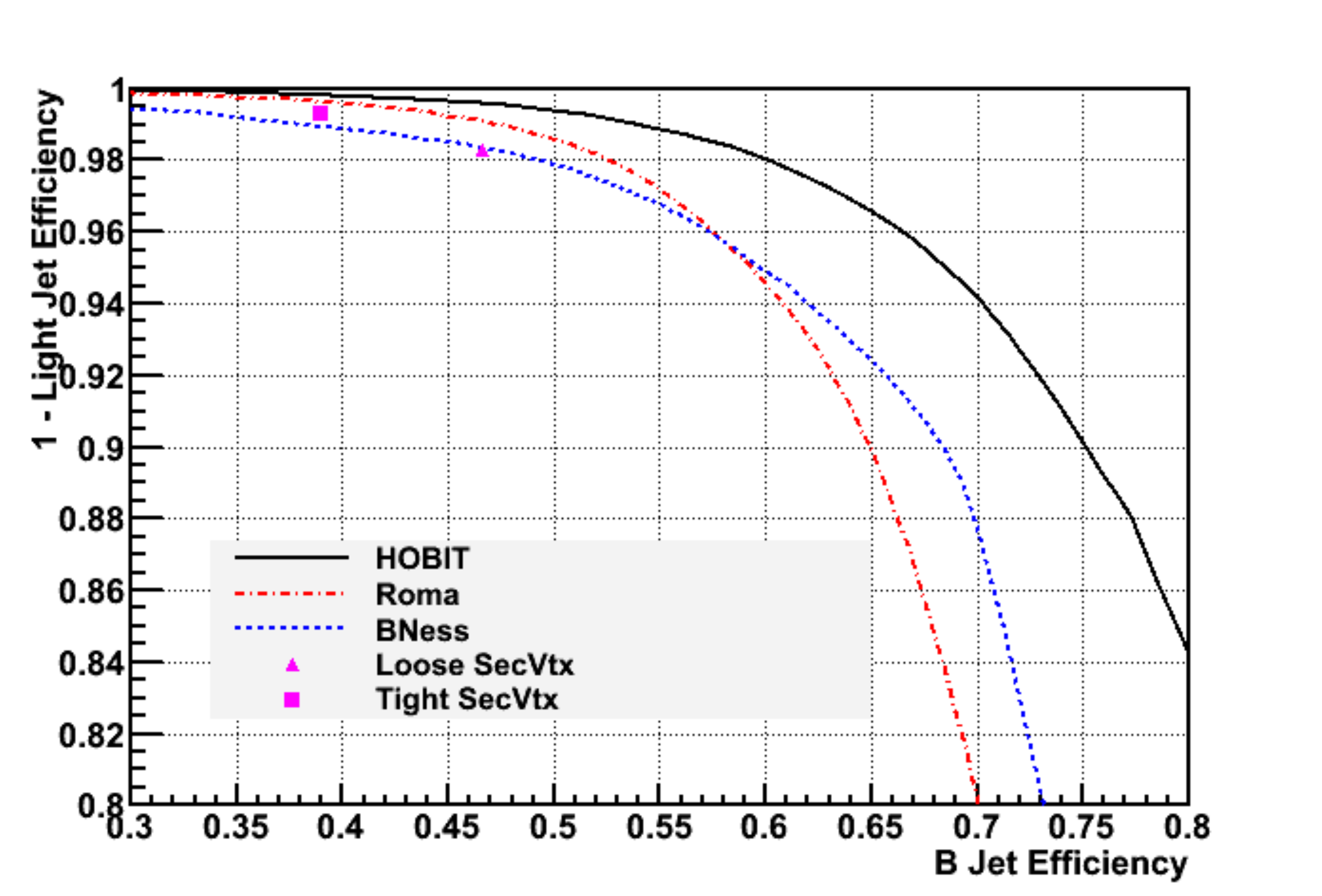}
    \caption{A comparison of the purity-efficiency tradeoffs for HOBIT versus RomaNN, Bness, and SecVtx loose and tight. A significant improvement over prior multivariate taggers is seen. 
             \label{roc_comp}}
  \end{center}
\end{figure}

The performance of a tagger is best evaluated by comparing its purity to tagging efficiency at  given operating points. We compare HOBIT's purity versus efficiency curve to the curves of the Bness and RomaNN taggers and to the purity versus efficiency performance of SecVtx at both its tight and loose operating points (Fig.~\ref{roc_comp}). Here, purity refers to the fraction of light-jets in $W+$jets MC which are not tagged as $b$-jets, and efficiency refers to the fraction of $b$ jets in light Higgs boson MC which are tagged. When evaluating tag efficiencies, the jets in both the numerator and denominator are required to have $E_T > $ 15 GeV and $|\eta| < 2$, the same $E_T$ and $\eta$ requirements as were placed on the jets in the training of HOBIT. Fig.~\ref{roc_comp} shows that for a given purity level, improvement in the absolute efficiency due to HOBIT is approximately 10\% over the Bness and RomaNN taggers, and approximately 15\% over the SecVtx tagger. 

\begin{figure}[htbp]
  \begin{center}
    \includegraphics[width=0.9\textwidth,clip=]{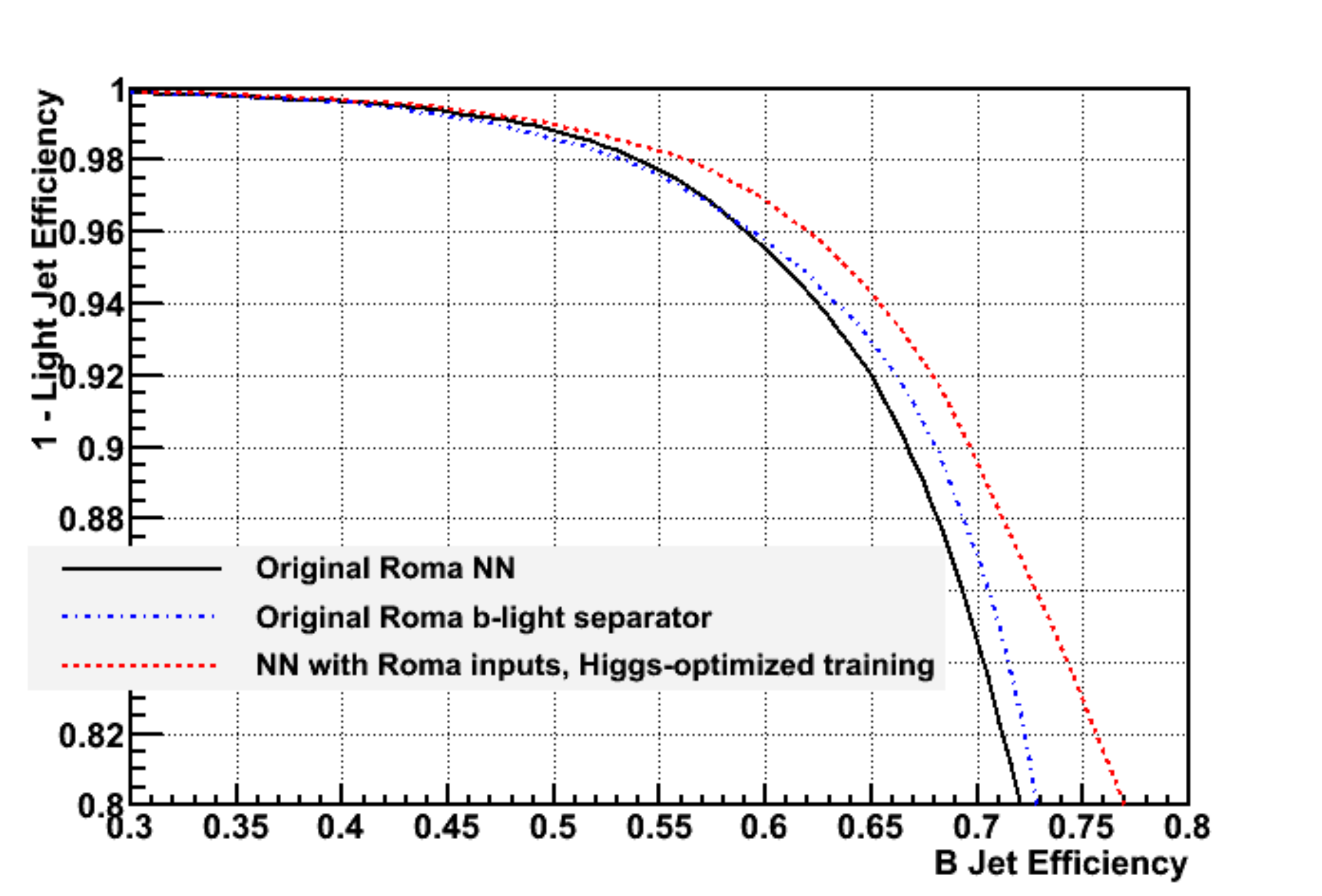}
    \caption{A comparison of the purity-efficiency tradeoffs for the original RomaNN tagger (as well as its $b$-light separator) and our version of the Higgs-optimized RomaNN tagger. 
             \label{roma_roc_comp}}
  \end{center}
\end{figure}

\begin{figure}[htbp]
  \begin{center}
    \includegraphics[width=0.9\textwidth,clip=]{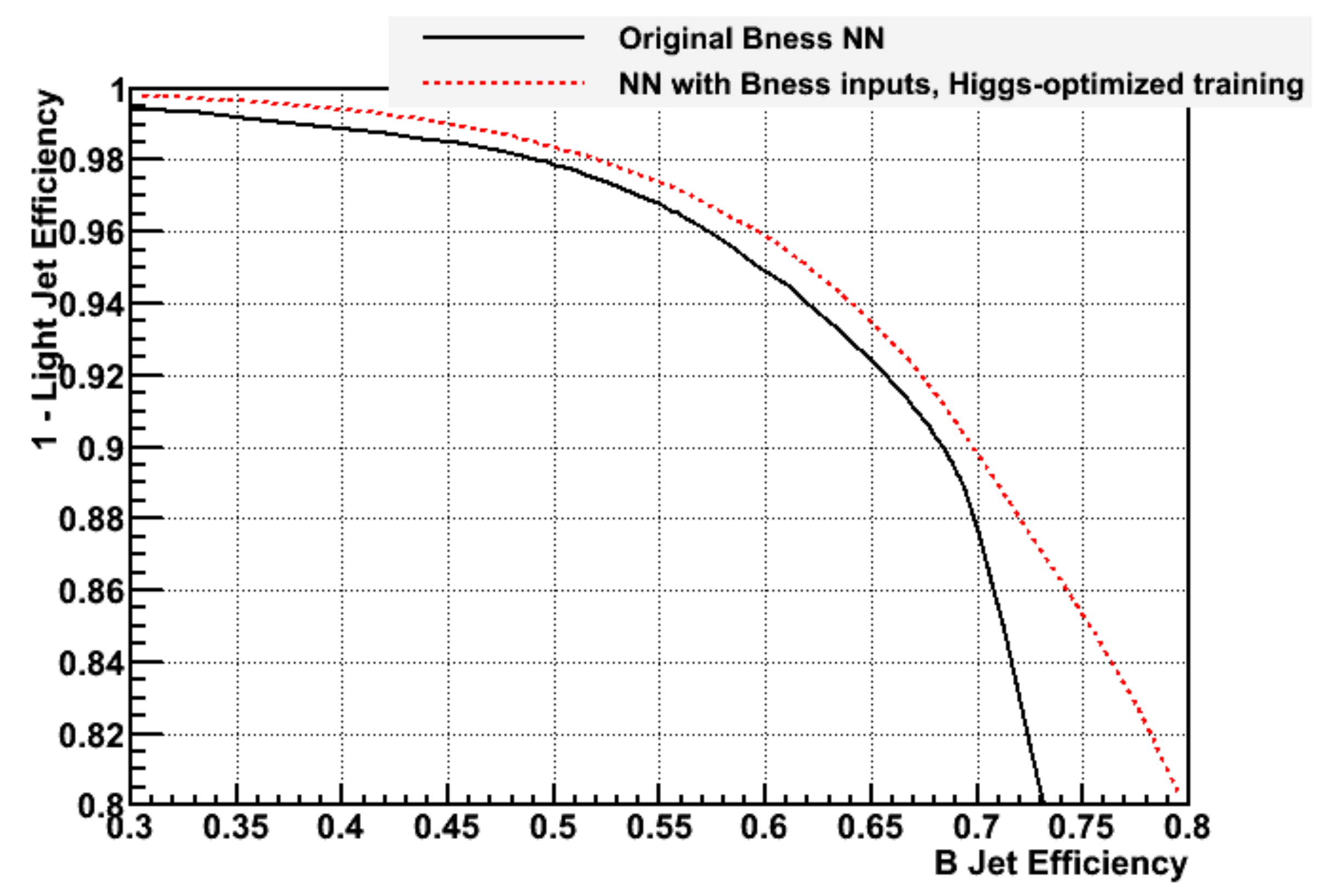}
    \caption{A comparison of the purity-efficiency tradeoffs for the original Bness tagger and our version of the Higgs-optimized Bness tagger. 
             \label{bness_roc_comp}}
  \end{center}
\end{figure}

We investigated how much of the improvement
	  in HOBIT over earlier taggers is due to the optimization on jets that
	  specifically originated from Higgs boson decays. To study this, we trained NN taggers
	  that take the same inputs as Bness and RomaNN using W$+$jets
	  and light Higgs boson MC, then compared the purity versus
	  efficiency curve with those of the original Bness and RomaNN
	  taggers, which were trained using $ZZ$ MC and $Z+$jets MC, respectively. The results can be seen in
	  Figs.~\ref{roma_roc_comp} and~\ref{bness_roc_comp}. In the case of the RomaNN comparison, not only is our
	  retrained RomaNN tagger compared with the original RomaNN
	  result, but also with RomaNN's $b$ versus light jet
	  separator. This is because the architecture of RomaNN
	  consisted of three different NN separators ($b$ versus light,
	  $b$ versus charm, light versus charm) which fed into the final
	  RomaNN separator. As we retrained using light and $b$ jets,
	  the comparison of the Higgs-optimized version of the RomaNN tagger with
	  the original $b$ versus light separator makes for a more fair
	  comparison. In both the Bness and RomaNN cases, the
	  improvement in absolute efficiency is approximately 2\%.

%%%%%%%%%%%%%%%%%%%%%%%%%%%%%%%%%%%%%%%%%%%%%%%%%%%%%%%%%%%%%%%%%%

% ======================================================================
\section{ Efficiency and Mistag Scale Factors }
\label{sec:scale_factors}

% 5. Efficiency and Mistag Scale factors
%         a. tt-bar cross section methodology
%         b. electron conversion methodology
%         c. systematic uncertainty discussion
% ======================================================================

In order to be used in a physics analysis, the performance of the
HOBIT $b$ tagger must be calibrated.  Historically, MC
modeling of $b$-tag efficiencies and mistag rates has not been
sufficient to use the uncorrected predictions of the MC. Instead, we use various techniques to measure the $b$-tagging efficiency and the mistag rate using CDF data. Examples of such techniques applied to the SecVtx algorithm are using jets containing electrons (therefore HF-enriched) for measuring the $b$-tagging efficiency~\cite{secvtxbtagsfelectron}, and using the rate at which jets have a displaced vertex reconstructed behind the primary vertex (``negative tags'') to estimate mistags~\cite{secvtxmistag}. For the tight
SecVtx tagger, the $b$-tag efficiency is found to be well predicted by the MC up to a scale factor (SF), where SF $= 0.96\pm 0.05$ for the full CDF dataset.
% Commented out 2/26 by John; charm SF isn't of direct interest in the HOBIT (b-vs-light) tagger
%We apply the same scale factor to the
%charm jets modeling in the MC, and double the systematic uncertainty. 
In order to utilize HOBIT to predict yields in data from MC simulation, a similar level of uncertainty in HOBIT's SF to that of SecVtx's SF is needed for each operating point.

An important difference between SecVtx and HOBIT is the absence of negative tags in
HOBIT, meaning the SecVtx mistag calculation technique cannot be applied.  Instead, we use two new 
techniques described below for calibrating $b$-tag SFs and providing mistag rates:
the ``$t{\bar{t}}$ cross section method'', and the ``electron conversion method''.

%----------------------------------------------------------------------------------
\subsection{Scale factors using the $t{\bar{t}}$ cross section method}
\label{sec:ttsfm}
The $t{\bar{t}}$ cross section method seeks to calibrate the predicted $b$-tagging
efficiency and the mistag rate in MC to match those measured in data
using $t{\bar{t}}$ candidate events in a $W+$3-or-more-jets sample under the
assumption that the $t{\bar{t}}$ cross section is known. The method is based upon
a previous analysis~\cite{nazimthesis} that simultaneously measured the SecVtx
$b$-tag SFs and the $t{\bar{t}}$ cross section. In that measurement, the rates of
singly and double tagged events provide a constraint which allows the
measurement of two unknowns. A two-dimensional fit was performed to
maximize the likelihood of observing the data counts as functions of the
SecVtx $b$-tag SF and the $t{\bar{t}}$ cross section.

This method has been repurposed such that the $t{\bar{t}}$ cross section is now
an input assumption, allowing for the calibration of the HOBIT $b$-tag
efficiency and the HOBIT mistag rate. We parameterize the resulting tag
rate in the MC samples as a 5-dimensional matrix, where each element is
the measured rate within a bin of the following five variables:  jet $E_T$,
jet $\eta$, the number of tracks in the jet, the number of primary vertices in the event,
and the $z$ location of the primary vertex from which the jet is calculated
to have originated. The matrix is similar to the SecVtx mistag matrix
\cite{secvtxmistag}, although of a lower dimension; the variables it has in common
with the SecVtx mistag matrix have the same binning between the two
matrices. For eight different HOBIT operating points, separate matrices are constructed for $b$, charm, and light jets.

The $W+$3-or-more-jets sample has an insufficient number of mistags to
calibrate the mistag SF, so we add a $W+$1 jet sample, which before $b$-tagging requirements is almost
pure $W+$light flavor (LF) events. After $b$ tagging, the $W+$1 jet sample consists of 
comparably sized $Wb{\bar{b}}$, $Wc{\bar{c}}$, $Wcj$, and mistagged $W+$LF events. The background predictions~\cite{SecVtx} involve scaling the total $W+$jets rate to data and subtracting off the non-$W+$jets components. The prediction of the
$W+$HF component of $W+$jets relies on the HF $K$-factor. This scaling adjusts leading-order theoretical predictions of the fraction of HF in $W+$jets events to account for higher-order corrections. We find that the $W+1$-jet data provides an independent handle on the mistag SF while the $b$-tag SF is constrained by the events with three or more jets. However, the dependence on the HF $K$-factor introduces a systematic
uncertainty that strongly affects the mistag SF.  For low values of the HOBIT cut, the mistag
rate is relatively high, and the relative contribution to the tagged $W$+1-jet sample from $W$+HF events is lower. This translates to a systematic uncertainty on the mistag SF due to the uncertainty on the HF $K$-factor that is lower at low HOBIT output values than at high HOBIT output values.

The maximum of the 2-d likelihood for the $b$-tag SF and the mistag SF is calculated given the observed data and fixed values of the HF $K$-factor, the $t{\bar{t}}$ cross section, and the minimum HOBIT output value. The dependence on the HF $K$-factor and the $t{\bar{t}}$ cross section are then
taken as sources of systematic uncertainty.  We assume $\sigma_{t{\bar{t}}}=7.04\pm 0.704 $~pb~\cite{higgsxsrec},
and take the HF $K$-factor to be $1.4\pm 0.4$.  

\begin{figure}
\includegraphics[width=\columnwidth]{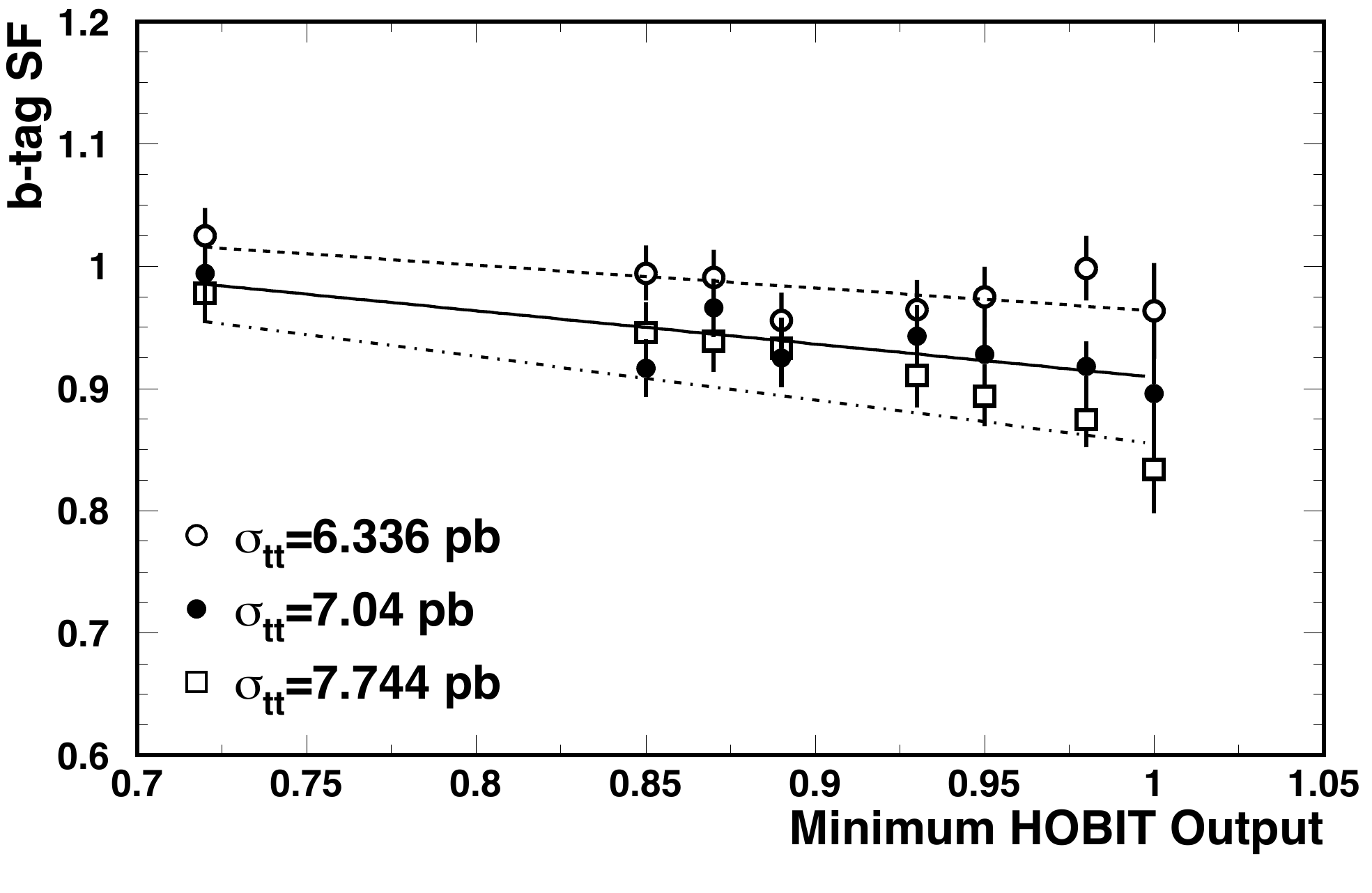}
\caption
{The measured value of the $b$-tag scale factor for the HOBIT tagger as a function of the minimum HOBIT output value.
Variations are shown assuming two values of the $t{\bar{t}}$ cross section.The straight lines are fits to the SFs
     assuming the central value of the $t\bar{t}$ cross section, and 
   $\sigma_{t{\bar{t}}} = 6.336$ pb, the more conservative case for the purpose of
     estimating uncertainties. The latter fit has been reflected through
     the central line to obtain a symmetric uncertainty band. }
\label{fig:ttbtsf}
\end{figure}

\begin{figure}
\includegraphics[width=\columnwidth]{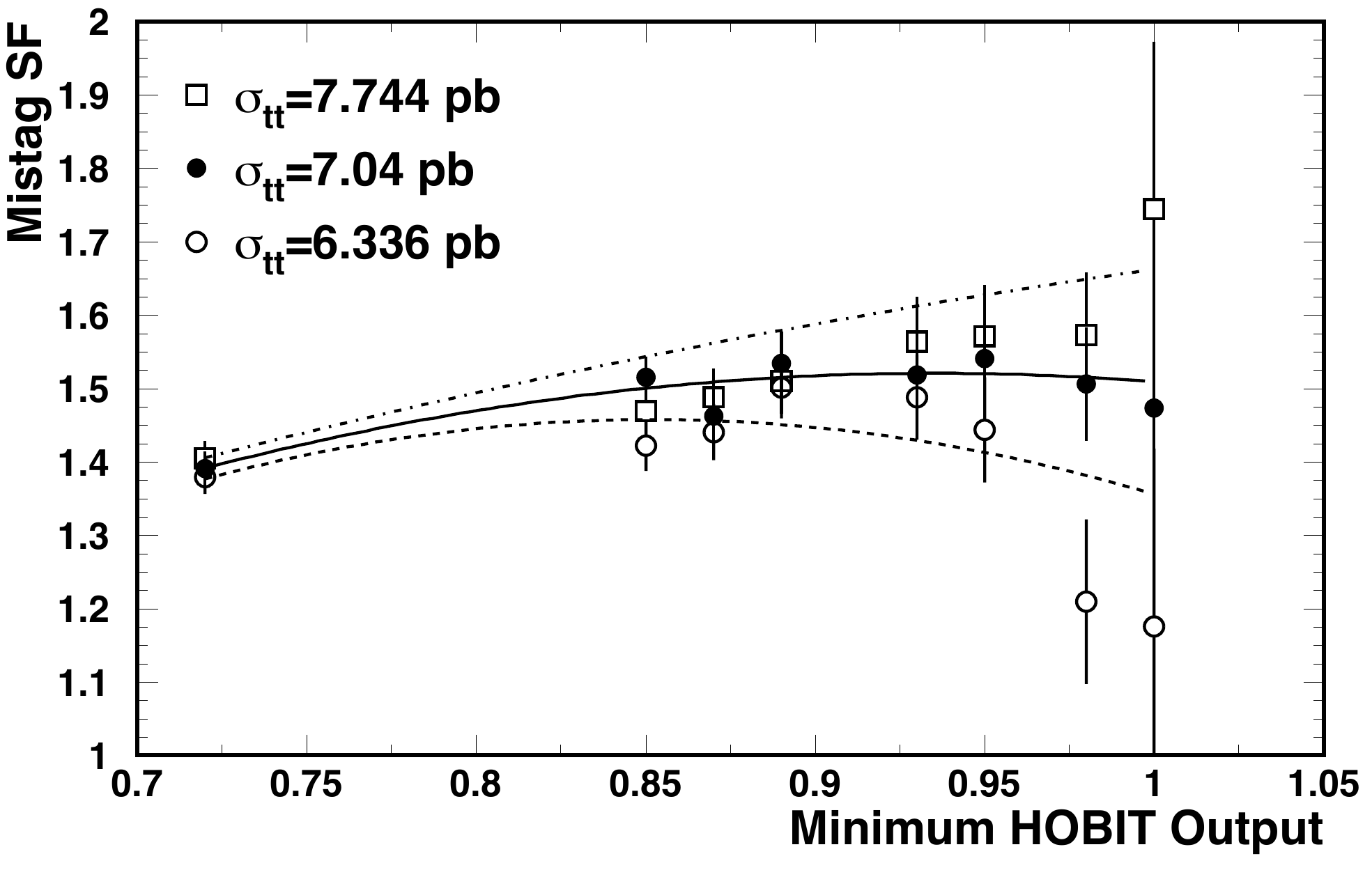}
\caption{The measured value of the mistag scale factor for the HOBIT tagger as a function of the minimum HOBIT output value.
Variations are shown assuming two values of the $t{\bar{t}}$ cross section. Parabolas are fit to the results assuming the central value of the
     $t{\bar{t}}$ cross section, and for $\sigma_{t{\bar{t}}} = 6.336$ pb. The latter has been reflected through the curve for the central value to obtain the depicted uncertainty band. }

%Parabolas are fit to the
%central values and the $\sigma_{t{\bar{t}}}=6.336$ pb values, and reflected over the central curve to obtain the other variation
%in order to be conservative.

\label{fig:ttmsf}
\end{figure}

The fitted $b$-tag and mistag SFs are shown in
Figures~\ref{fig:ttbtsf} and~\ref{fig:ttmsf}, respectively, as functions of the minimum HOBIT output value. The curves represent a linear fit to the $b$-tag SF as a function of the minimum HOBIT output value, and a parabolic fit to the mistag SF. The variation due to $\sigma_{t{\bar{t}}}$ is also shown, where we take the larger of the two shifts in the result due to an increase/decrease in $\sigma_{t{\bar{t}}}$ and then symmetrize the uncertainty. 

The determination of the $b$-tag and mistag SFs are subject to the same sources of systematic
uncertainty as a measurement of $\sigma_{t{\bar{t}}}$~\cite{ttxsz}.  Specifically,
the $t{\bar{t}}$ acceptance depends on initial-state radiation and final-state radiation (ISR+FSR), parton distribution functions (PDFs), jet energy scale, trigger efficiencies and lepton identification efficiencies. The luminosity uncertainty, although nearly absent in the results of Ref.~\cite{ttxsz}, also contributes to the overall systematic uncertainty.  

%% All sources of systematic uncertainty have the same impact on this
%% method as changing the assumed  $\sigma_{t{\bar{t}}}$.  Thus the dependence of the measured $b$-tag
%% and mistag scale factors on the assumed value of $\sigma_{t{\bar{t}}}$ provides the mechanism by which
%% the other sources of systematic uncertainty can be evaluated.

\begin{figure}
\includegraphics[width=\columnwidth]{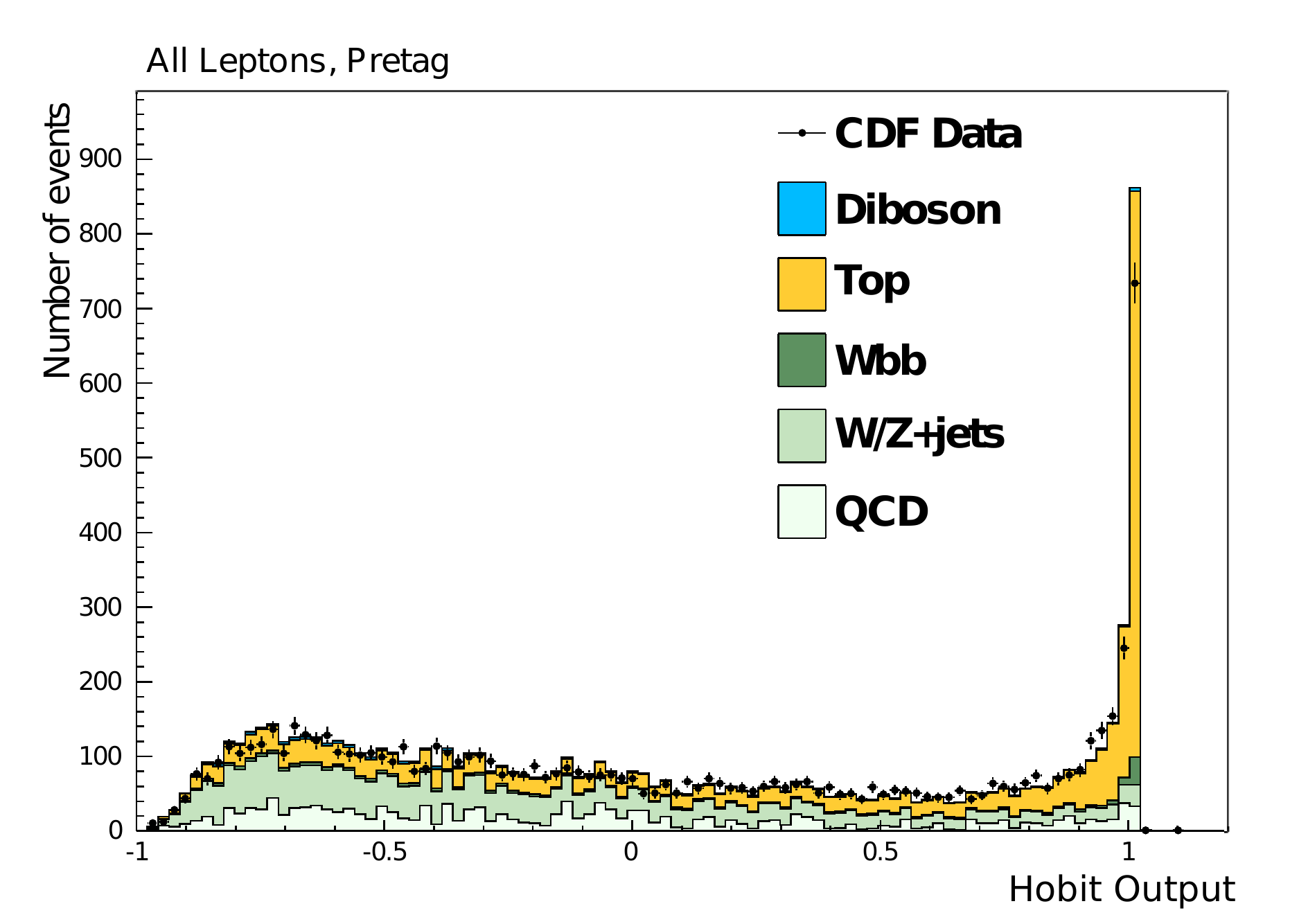}
\caption{Data versus\ MC, the HOBIT output distribution of the highest $E_T$ jet from events in the $WH \rightarrow l\nu b \bar{b}$ sample before a requirement of a $b$-jet tag.}

\label{fig:WH_hobit}
\end{figure}

\begin{figure}
\includegraphics[width=\columnwidth]{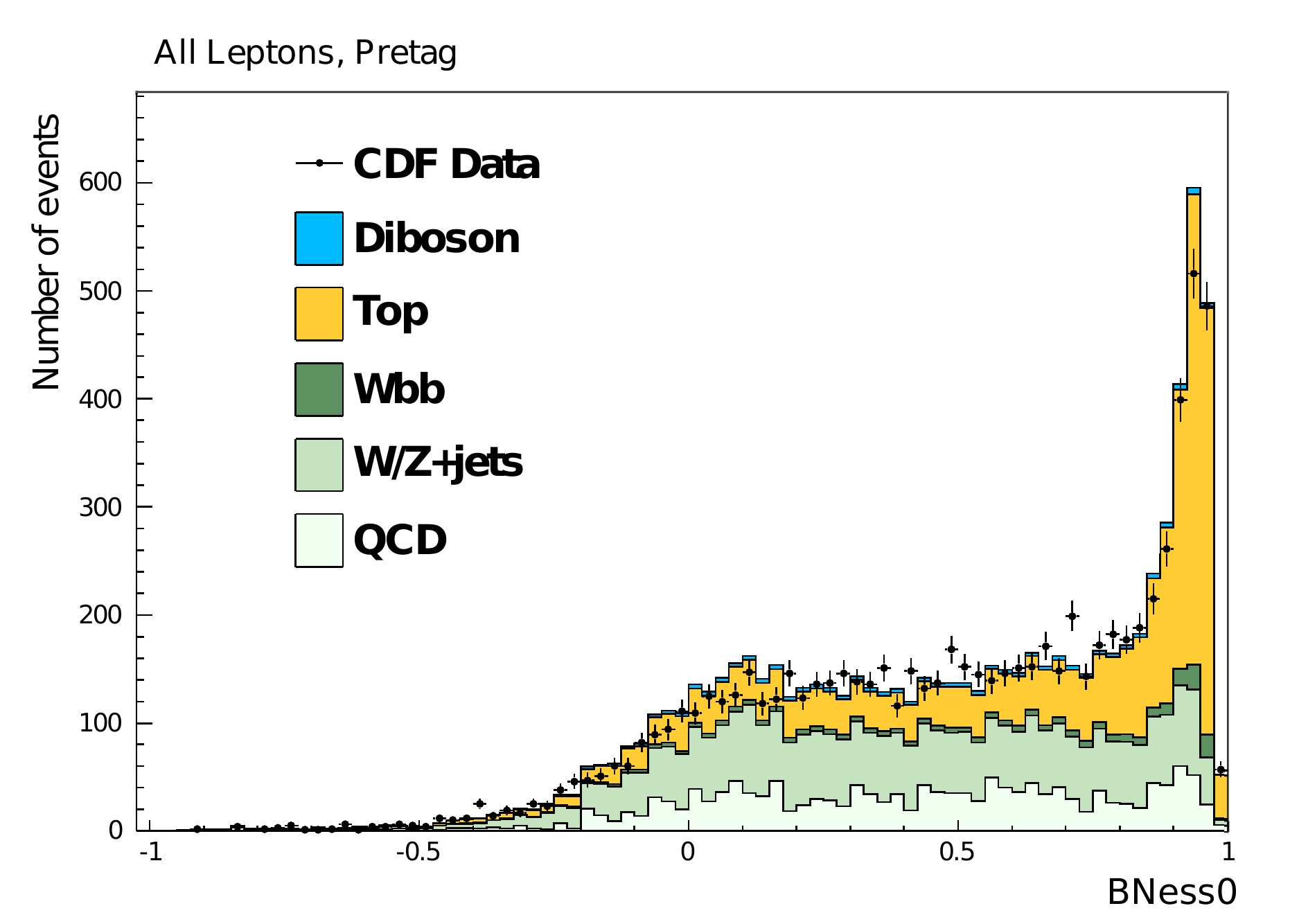}
\caption{Data versus\ MC, highest track Bness of the highest $E_T$ jet from events in the $WH \rightarrow l\nu b \bar{b}$ sample before a requirement of a $b$-jet tag.}

\label{fig:WH_bness0}
\end{figure}

\begin{figure}
\includegraphics[width=\columnwidth]{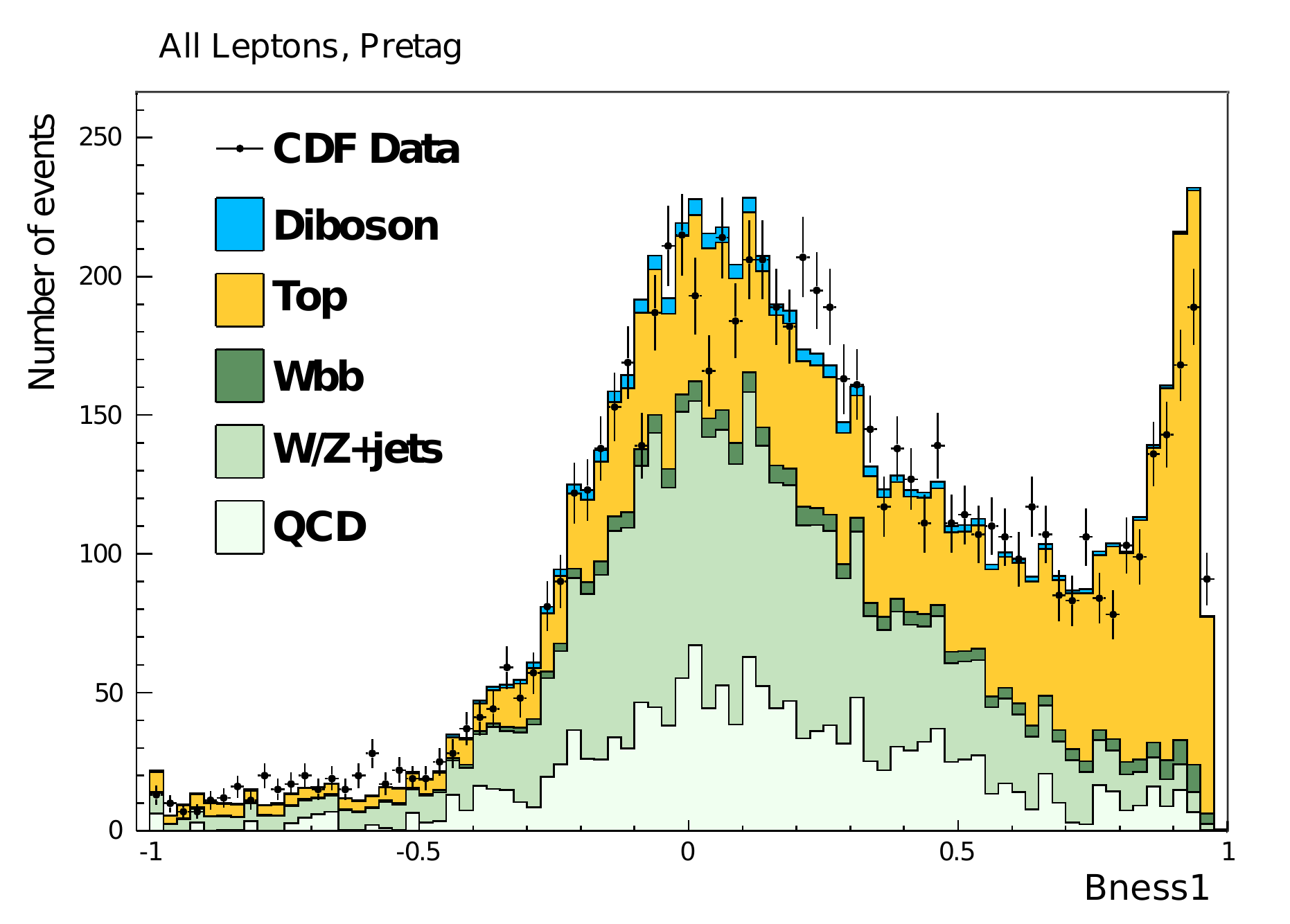}
\caption{Data versus\ MC, second highest track Bness of the highest $E_T$ jet from events in the $WH \rightarrow l\nu b \bar{b}$ sample before a requirement of a $b$-jet tag.}

\label{fig:WH_bness1}
\end{figure}

\begin{figure}
\includegraphics[width=\columnwidth]{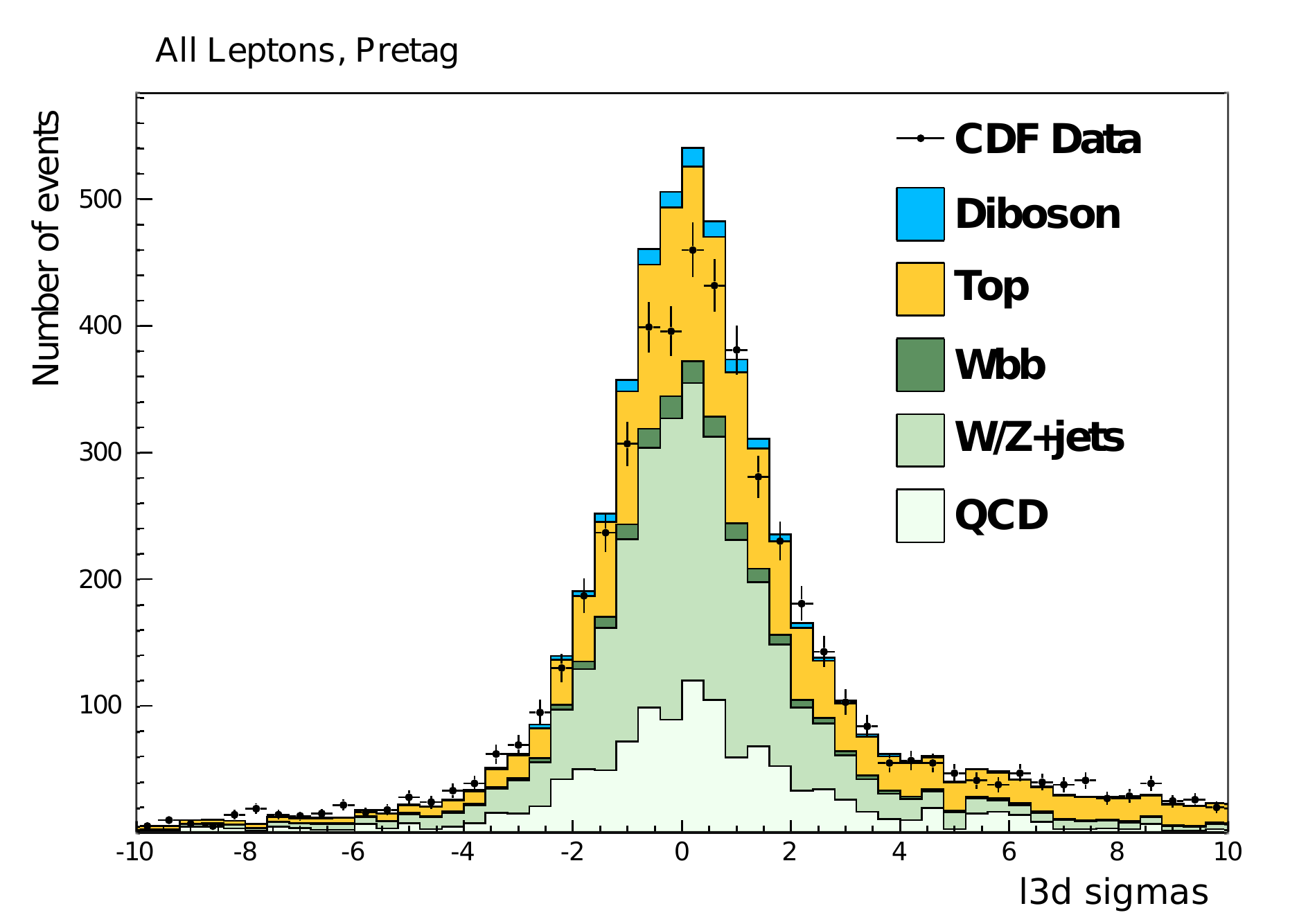}
\caption{Data versus\ MC, 3-d displacement significance of most HF-like displaced vertex of the highest $E_T$ jet from events in the $WH \rightarrow l\nu b \bar{b}$ sample before a requirement of a $b$-jet tag.}

\label{fig:WH_l3dsig}
\end{figure}

\begin{figure}
\includegraphics[width=\columnwidth]{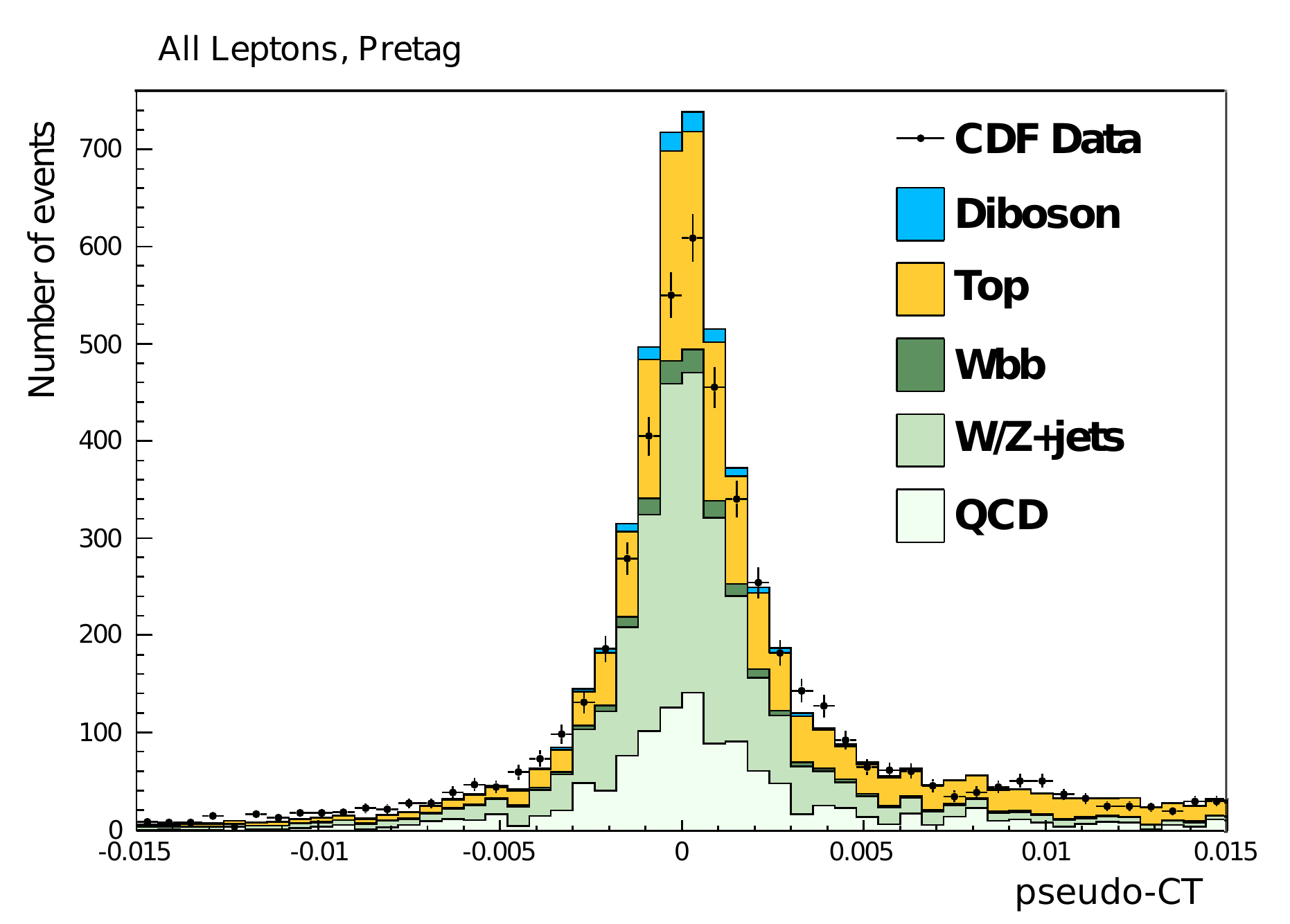}
\caption{Data versus\ MC, pseudo-c$\tau$ of most HF-like displaced vertex of the highest $E_T$ jet from events in the $WH \rightarrow l\nu b \bar{b}$ sample before a requirement of a $b$-jet tag.}

\label{fig:WH_pseudoctau}
\end{figure}

For the loose (0.72) and tight (0.98) HOBIT operating points, this method yields efficiency SFs of 0.997 $\pm$ 0.037 and 0.917 $\pm$ 0.069, respectively. The mistag rate SFs are $1.391 \pm 0.202$ and $1.515 \pm 0.291$. A complete table of systematic uncertainties for the efficiency SF is shown in Table~\ref{tab:beff_sf_uncertainty_ttbar}, and for the mistag matrix SF in Table~\ref{tab:mistag_sf_uncertainty_ttbar}. Figures~\ref{fig:WH_hobit},~\ref{fig:WH_bness0},~\ref{fig:WH_bness1},~\ref{fig:WH_l3dsig}, and~\ref{fig:WH_pseudoctau} show validation plots comparing properties of the highest $E_T$ jet (HOBIT output, and select HOBIT inputs) in $WH \rightarrow l\nu b \bar{b}$ candidate events before any $b$-tag requirements or SF corrections are applied for MC versus data. Good agreement is seen between MC and data.

\begin{table}[htdp]
\caption{The systematic uncertainties for the $b$-jet tagging efficiency scale factor from the $\sigma(t\bar{t})$ method measurement.  This uncertainty must be combined with the electron method scale factor uncertainty; the two should be treated as uncorrelated. The uncertainties shown below are absolute shifts.}
\begin{center}
\begin{tabular}{|c|c|c|c|} \hline
\multicolumn{2}{|c|}{b-eff SF $\sigma (t\bar{t})$ method}&\multicolumn{2}{|c|}{HOBIT Operating Point}\\ \hline
\multicolumn{2}{|c|}{source} &Loose&Tight\\ \hline
\multirow{2}{*}{$\sigma (t\bar{t})$}& up & -0.011 &-0.019\\
& down& 0.011&0.019\\ \hline
 \multirow{2}{*}{luminosity }& up & -0.004 &-0.055\\
& down & 0.007& 0.012\\ \hline
 \multirow{2}{*}{jet energy scale }& up & -0.005  &-0.007 \\
& down & 0.005  &0.007 \\ \hline
 \multirow{2}{*}{generator}& up & 0.003& 0.005\\
& down &  -0.003& -0.005\\ \hline
 \multirow{2}{*}{ISR/FSR}& up & -0.001& -0.001\\
& down &  0.001&  0.001\\ \hline
 \multirow{2}{*}{$t \rightarrow Wb$ branching ratio}& up & -0.001& -0.001\\
& down &  0.001&  0.001\\ \hline
 \multirow{2}{*}{Trigger}& up & -0.001& -0.001\\
& down &  0.001&  0.001\\ \hline
 \multirow{2}{*}{PDF}& up & 0.001&  0.001\\
& down &  -0.001& -0.001\\ \hline
 \multirow{2}{*}{W+j kfactor}& up & 0.009 & 0.006\\
& down &  -0.009 & -0.006\\ \hline
 \multirow{2}{*}{Statistics}& up & 0.014 & 0.008\\
& down & -0.014 & -0.008\\ \hline \hline
\multirow{2}{*}{total} & up & 0.022&0.026\\
& down & -0.022&-0.026\\\hline
\end{tabular}
\end{center}
\label{tab:beff_sf_uncertainty_ttbar}
\end{table}%

\begin{table}[htdp]
\caption{The systematic uncertainties for the mistag rate scale factor from the $\sigma(t\bar{t})$ method measurement.  This uncertainty must be combined with the electron method scale factor uncertainty; the two should be treated as uncorrelated. The uncertainties shown below are absolute shifts.}
\begin{center}
\begin{tabular}{|c|c|c|c|} \hline
\multicolumn{2}{|c|}{mistag SF $\sigma (t\bar{t})$ method}&\multicolumn{2}{|c|}{HOBIT Operating Point}\\ \hline
\multicolumn{2}{|c|}{source} &Loose&Tight\\ \hline
\multirow{2}{*}{$\sigma (t\bar{t})$}& up & 0.007 &0.090\\
& down&  -0.007 &-0.090\\ \hline
 \multirow{2}{*}{luminosity }& up & 0.004 &0.055\\
& down &  -0.004 &-0.055\\ \hline
 \multirow{2}{*}{jet energy scale }& up & 0.003 &0.037\\
& down &  -0.003  &-0.037\\ \hline
 \multirow{2}{*}{generator}& up & 0.002& 0.023\\
& down & -0.002& -0.023\\ \hline
 \multirow{2}{*}{ISR/FSR}& up & 0.000& 0.005\\
& down &  -0.000& -0.005\\ \hline
 \multirow{2}{*}{$t \rightarrow Wb$ branching ratio}& up & 0.000& 0.005\\
& down & -0.000& -0.005\\ \hline
 \multirow{2}{*}{Trigger}& up & 0.000& 0.005\\
& down & -0.000& -0.005\\ \hline
 \multirow{2}{*}{PDF}& up &  0.000& 0.005\\
& down & -0.000& -0.005\\ \hline
 \multirow{2}{*}{W+j kfactor}& up & -0.091 & -0.135\\
& down &  0.055 & 0.081\\ \hline
 \multirow{2}{*}{Statistics}& up & 0.024 & 0.125\\
& down & -0.024 & -0.125\\ \hline \hline
\multirow{2}{*}{total} & up & 0.094&0.217\\
& down &  -0.060&-0.180\\ \hline
\end{tabular}
\end{center}
\label{tab:mistag_sf_uncertainty_ttbar}
\end{table}%

%--------------------------------------------------------------------------
\subsection{ Scale factors using the electron conversion method }

A second method of calculating the correction for the HOBIT MC response involves a modification of the traditional SecVtx efficiency SF algorithm in a way that does not require the concept of
a ``negative tag''~\cite{secvtxbtagsfelectron}. However, like the SecVtx technique, this method takes advantage of the HF
enhancement among jets containing electrons, discriminating between HF and LF jets based upon whether the electron is identified as coming from a photon conversion.

The event sample consists of back-to-back dijet events where one jet contains an electron candidate (the electron jet, or ``e-jet''), while its opposite jet has no such requirement (the away jet, or ``a-jet''). We can label each jet originating
either from an HF quark (``B'') or a light flavor quark or gluon (``Q'') and
categorize each event as $N_{XY}$, where the e-jet has flavor X and the a-jet has
flavor Y. Then the total number of events ($N^e$) is $$N^e = N_{BB} + N_{BQ} + N_{QB} + N_{QQ}$$ 
and the HF fraction of the e-jets is 
  $$ F_B = (N_{BB}+N_{BQ})/N^e.$$ 
Applying a $b$ tag on the e-jet with a tagging efficiency ($\epsilon^e$) and a mistag rate 
($\epsilon_{mis}$), the number of $b$-tagged e-jets ($N^e_+$) is 
  $$N^e_+ = \epsilon^e \cdot (N_{BB}+N_{BQ}) + \epsilon^e_{mis}\cdot (N_{QB} + N_{QQ}) .$$ 

Assuming the fraction of light flavor jets with conversions is $f^c$ and the conversion finding efficiency is $\epsilon^c$ for the light flavor jets and $\epsilon^0$ for the HF 
jets, we can obtain the number of e-jets identified from the conversion $N^{ec}$ as
  $$ N^{ec} = \epsilon^0 \cdot (N_{BB}+N_{BQ}) + \epsilon^c \cdot f^c \cdot (N_{QB}+N_{QQ}) $$
After tagging, the number of $b$-tagged conversion e-jets ($N^{ec}_+$) becomes
  $$ N^{ec}_+ = k\cdot \epsilon^e \cdot \epsilon^0 \cdot (N_{BB}+N_{BQ}) + \epsilon^e_{mis}\cdot \epsilon^c \cdot f^c \cdot (N_{QB}+N_{QQ}) ,$$ 
where $k$ is the ratio of the $b$-tag efficiency for an HF e-jet identified as a conversion to that for one that is not. 

The previous two equations allow us to solve for $\epsilon_{mis}$ and $\epsilon^e$: 
   $$\epsilon_{mis} = (N^{ec}_+-k\cdot \epsilon^0 \cdot N^e_+)/(N^{ec} - \epsilon^0 \cdot N^e\cdot(k+(1-k)\cdot F_B)) $$ 
and
   $$\epsilon^e = (N^e_+ -\epsilon_{mis} \cdot N^e \cdot (1-F_B))/(N^e\cdot F_B). $$  
Here, all terms that are not the mistag and efficiency rates can be counted
directly in data, taken from MC ($k$), measured in data ($F_B$) or both taken from MC
and/or measured in data ($\epsilon^0$). In the case of $F_B$, we can simply use the traditional SecVtx electron method~\cite{secvtxbtagsfelectron} to give us this value. For $\epsilon^0$, obtaining this quantity from MC is trivial, as we have truth information available. To calculate it from data, we look at the rate at which positively SecVtx-tagged jets are found to contain conversion electrons and then adjust this rate using negatively-SecVtx-tagged jets. 

The resulting tagging efficiency SFs for the loose and tight HOBIT outputs are 0.986 $\pm$ 0.066 and 0.949 $\pm$ 0.044 respectively, in good agreement with the results from the $t\bar{t}$ method. Some of the largest contributors to the systematic component of these uncertainties includes the difference between the results when we use the MC-calculated $\epsilon^0$ versus the data-calculated version and the fact that $b$-jets containing electrons tend to leave fewer tracks than typical $b$-jets. 

\begin{figure}[htbp]
  \begin{center}
    \includegraphics[width=0.9\textwidth,clip=]{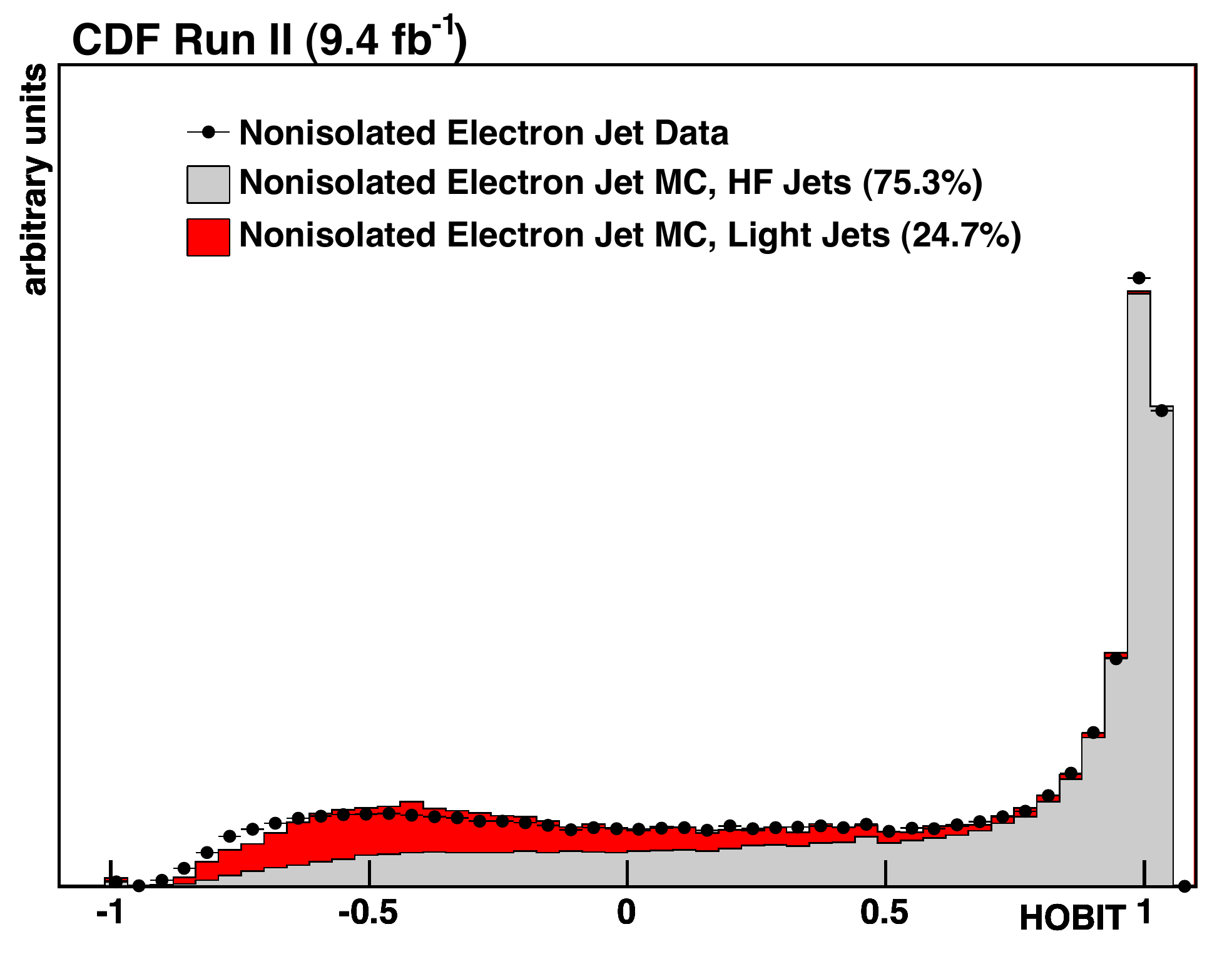}
    \caption{HOBIT output for electron jets, data versus MC. Relative proportions of HF to light jets are determined via a fit of the two MC templates to the data. 
             \label{fig:ejetHOBIT}}
  \end{center}
\end{figure}

\begin{figure}[htbp]
  \begin{center}
    \includegraphics[width=0.9\textwidth,clip=]{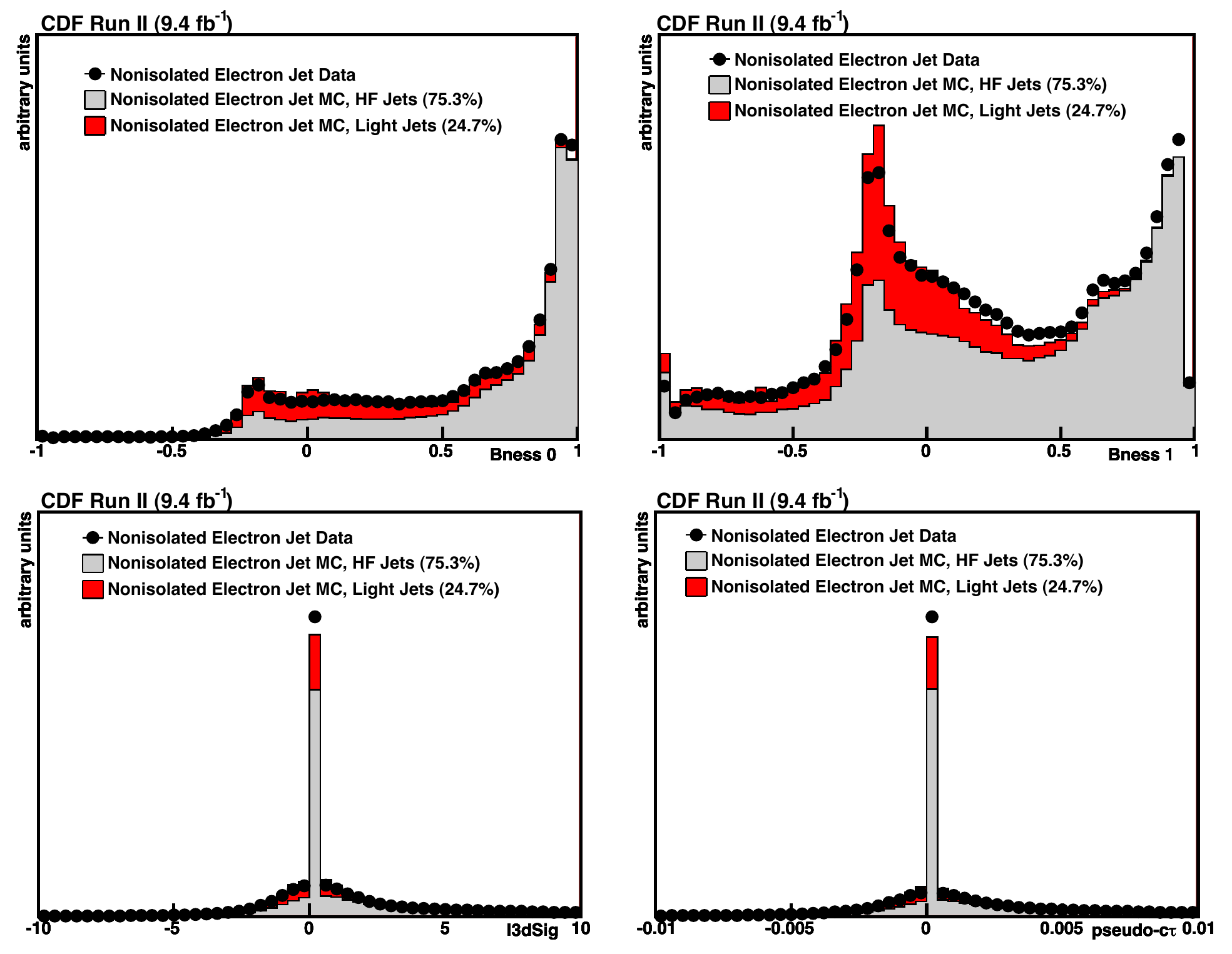}
    \caption{Comparison of select HOBIT inputs for electron jets, data versus MC. 
             \label{fig:ejetinputs}}
  \end{center}
\end{figure}

The SFs on the mistag rate for the loose and tight HOBIT operating points are 1.28 $\pm$ 0.17 and 1.42 $\pm$ 0.89, respectively, also consistent with the results of the $t\bar{t}$ method. As a check, we compare e-jets in data and MC (Figs.~\ref{fig:ejetHOBIT} and~\ref{fig:ejetinputs}), after purifying the HF content by requiring the away jet to be tight SecVtx tagged and the electron in the e-jet to not be identified as a conversion. The fraction of HF versus light jet MC used in these plots is determined via a fit of MC templates to the HOBIT distribution in data. 

\begin{table}[htdp]
\caption{The systematic uncertainties for the $b$-jet tagging efficiency scale factor from the electron method measurement.  This uncertainty must be combined with the $\sigma(t\bar{t})$ method scale factor uncertainty; the two should be treated as uncorrelated. The uncertainties shown below are absolute shifts.}
\begin{center}
\begin{tabular}{|c|c|c|c|} \hline
\multicolumn{2}{|c|}{b-eff SF electron method}&\multicolumn{2}{|c|}{HOBIT Operating Point}\\ \hline
\multicolumn{2}{|c|}{source} &Loose&Tight\\ \hline
 \multirow{2}{*}{over eff. }& up & 0.009 & 0.014\\
& down & -0.009 & -0.014\\\hline
 \multirow{2}{*}{prescale coor.}& up & 0.001 &0.011\\
& down &  -0.001  &-0.011\\\hline
 \multirow{2}{*}{Et depend.}& up & 0.010& 0.003\\
& down & -0.010& -0.003\\\hline
 \multirow{2}{*}{semi-lep bias}& up & 0.010& 0.006\\
& down &-0.010& -0.006\\\hline
 \multirow{2}{*}{charm model}& up & 0.001&  0.002\\
& down &  -0.001&  -0.002\\ \hline
 \multirow{2}{*}{Stats}& up & 0.016 & 0.018\\
& down & -0.016 & -0.018\\ \hline \hline
\multirow{2}{*}{total} & up & 0.023&0.026\\
& down & -0.023&-0.026\\\hline
\end{tabular}
\end{center}
\label{tab:beff_sf_uncertainty_eletron}
\end{table}%

\begin{table}[htdp]
\caption{The systematic uncertainties for the mistag rate scale factor from the electron method measurement.  This uncertainty must be combined with the $\sigma(t\bar{t})$ method scale factor uncertainty; the two should be treated as uncorrelated. The uncertainties shown below are absolute shifts.}
\begin{center}
\begin{tabular}{|c|c|c|c|} \hline
\multicolumn{2}{|c|}{b-eff SF electron method}&\multicolumn{2}{|c|}{HOBIT Operating Point}\\ \hline
\multicolumn{2}{|c|}{source} &Loose&Tight\\ \hline
 \multirow{2}{*}{over eff. }& up & 0.024 & 0.092\\
& down & -0.024 & -0.092\\ \hline
 \multirow{2}{*}{prescale coor.}& up & 0.010 &0.003\\
& down &  -0.010 &-0.003\\ \hline
 \multirow{2}{*}{Et depend.}& up & 0.014& 0.018\\
& down &  -0.014& -0.018\\ \hline
 \multirow{2}{*}{semi-lep bias}& up & 0.040& 0.055\\
& down & -0.040& -0.055\\ \hline
 \multirow{2}{*}{charm model}& up & 0.001&  0.004\\
& down &-0.001&  -0.004\\ \hline
 \multirow{2}{*}{Stats}& up & 0.078 & 0.163\\
& down & -0.078 & -0.163\\ \hline\hline
\multirow{2}{*}{total} & up & 0.092&0.196\\
& down &  -0.092&-0.196\\ \hline
\end{tabular}
\end{center}
\label{tab:mistag_sf_uncertainty_eletron}
\end{table}%

\subsection{SF Combination}

When combining the correction SFs for the MC $b$-tag efficiency from the electron and $t\bar{t}$ method, we obtain 0.993 $\pm$ 0.032 (for HOBIT's loose operating point, 0.72) and 0.937 $\pm$ 0.037 (HOBIT's tight operating point, 0.98). The combined results for the mistag rates are 1.331 $\pm$ 0.130 and 1.492 $\pm$ 0.277, respectively. Due to the uncertainties in the electron and $t\bar{t}$ methods being uncorrelated, the combination is straightforward. This results in a greater than 25\% reduction in the size of the uncertainty on the $b$-tag efficiency in comparison to the previous most widely used CDF b-tagging algorithm, SecVtx.

% ======================================================================
%\subsection{ Systematic uncertainties }

\section{Conclusion}
\label{sec:conclusion}

We have developed an NN-based $b$ identification tagger
which improves upon the best ideas of previous CDF taggers, has a very
generous taggability requirement, and has been optimized for $H
\rightarrow b\bar{b}$ searches, the primary decay channel of the light
Higgs boson at the Tevatron. Using two uncorrelated and innovative methods, we found tagging efficiencies, mistag rates, and data-to-MC scale factors that are in good agreement. The combination of these methods results in a greater than 25\% reduction in the $b$-tag efficiency uncertainty compared to SecVtx, the previous most widely used CDF $b$-tagging algorithm. In the current light Higgs boson analyses at CDF, we estimate that replacing previous tagging algorithms with HOBIT results in a 10-20\% improvement in Higgs boson sensitivity.

\section*{Acknowledgements}
The authors thank the CDF collaboration, the Fermilab staff and the technical staffs of the participating institutions for their vital contributions. This work was supported by the US Department of Energy and the Fermilab Research Alliance International Fellowship.

%%%%%%%%%%%%%%%%%%%%%%%%%%%%%%%%%%%%%%%%%%%%%%%%%%%%%%%%%%%%%%%%%%

\end{document}